\documentclass{pasa}

\usepackage{graphicx}
\usepackage[normalem]{ulem}

\def\cp{C$^+$}
\def\hh{H$_2$}
\def\hho{H$_2$O}
\def\coo{CO$_2$}
\def\cchh{C$_2$H$_2$}

\def\mic{$\mu$m}
\def\scm{cm$^{-2}$}
\def\pow#1#2{#1$\times$10$^{#2}$}
\def\msol{M$_{\odot}$}
\def\lsol{L$_{\odot}$}
\def\zsol{Z$_{\odot}$}

\def\spicas{SPICA\textsc{{\char13}s}}
\def\gtsim{{_>\atop{^\sim}}}

\DeclareRobustCommand{\ion}[2]{\textup{#1\,\textsc{\lowercase{#2}}}}

\newcommand{\fsl}[3]{[\ion{#1}{#2}] #3\,\mic}

\title{Probing the baryon cycle of galaxies with SPICA mid- and far-infrared observations}
\author[Van der Tak et al.]{F.F.S. van der Tak$^{1,2}$, 
S.C. Madden$^3$, 
P. Roelfsema$^1$;
L. Armus$^4$,
M. Baes$^5$,
J. Bernard-Salas$^6$,
A. Bolatto$^7$,
S. Bontemps$^8$, 
C. Bot$^9$,
C.M. Bradford$^4$,
J. Braine$^8$, 
L. Ciesla$^3$, 
D. Clements$^{10}$,
D. Cormier$^{11,3}$, 
J.A. Fern\'andez-Ontiveros$^{12,13,14}$,
F. Galliano$^3$, 
M. Giard$^{15}$,
H. Gomez$^{16}$,
E. Gonz\'alez-Alfonso$^{17}$,
F. Herpin$^8$,
D. Johnstone$^{18,19}$
A. Jones$^{20}$, 
H. Kaneda$^{21}$, 
F. Kemper$^{22}$,
V. Lebouteiller$^3$,
I. De Looze$^{23}$, 
M. Matsuura$^{16}$, 
T. Nakagawa$^{24}$,
T. Onaka$^{25}$,
P. P\'erez-Gonz\'alez$^{26}$,
R. Shipman$^1$,
L. Spinoglio$^{14}$}
\jid{PASA}
\doi{10.1017/pas.\the\year.xxx}
\jyear{\the\year}

\date{Submitted 20 July 2017; accepted 29 November 2017}

\usepackage[authoryear]{natbib}
\bibpunct{(}{)}{;}{a}{}{,}
\usepackage{aas_macros}
\usepackage{hyperref} 
\hypersetup{colorlinks,citecolor=blue,linkcolor=blue,urlcolor=blue}

\begin{document}%
\begin{abstract}
The SPICA mid and far-infrared telescope will address fundamental issues in our understanding of star formation and ISM physics in galaxies. 
A particular hallmark of SPICA is the outstanding sensitivity enabled by the cold telescope, optimized detectors, and wide instantaneous bandwidth throughout the mid- and far-infrared.
The spectroscopic, imaging and polarimetric observations that SPICA will be able to collect will help in clarifying the complex physical mechanisms which underlie the baryon cycle of galaxies. 
In particular:
(i) The access to a large suite of atomic and ionic fine-structure lines for large samples of galaxies will shed light on the origin of the observed spread in star formation rates within and between galaxies.
(ii) Observations of HD rotational lines (out to $\sim$10\,Mpc) and fine structure lines such as \fsl{C}{ii}{158}\ (out to $\sim$100\,Mpc) will clarify the main reservoirs of interstellar matter in galaxies, including phases where CO does not emit.
(iii) Far-infrared spectroscopy of dust and ice features will address uncertainties in the mass and composition of dust in galaxies, and the contributions of supernovae to the interstellar dust budget will be quantified by photometry and monitoring of supernova remnants in nearby galaxies.
(iv) Observations of far-infrared cooling lines such as \fsl{O}{i}{63}\ from star-forming molecular clouds in our Galaxy will evaluate the importance of shocks to dissipate turbulent energy.
The paper concludes with requirements for the telescope and instruments, and recommendations for the observing strategy.
\end{abstract}
\maketitle%

{\bf Preface}

\vspace{0.5cm}
\noindent
The following set of articles describe in detail the science goals of the future Space Infrared telescope for Cosmology and Astrophysics (SPICA). 
The SPICA satellite will employ a 2.5m telescope, actively cooled to below 8\,K, and a suite of mid- to far-IR spectrometers and photometric cameras, equipped with state of the art detectors. 
In particular, the SPICA Far Infrared Instrument (SAFARI) will be a grating spectrograph with low ($R=300$) and medium ($R\sim3000-11000$) resolution observing modes instantaneously covering the $35-230$ $\mu$m wavelength range. 
The SPICA Mid-Infrared Instrument (SMI) will have three operating modes:  a large field of view ($12'\times10'$) low-resolution $17-36$ $\mu$m spectroscopic ($R\sim50-120$) and photometric camera at 34 $\mu$m, a medium resolution ($R\sim2000$) grating spectrometer covering wavelengths of $18-36$ $\mu$m and a high-resolution echelle module ($R\sim28000$) for the $12-18$ $\mu$m domain.
A  large field of view ($80''\times80''$), three channel (110 $\mu$m, 220 $\mu$m and 350 $\mu$m) polarimetric camera (POL) will also be part of the instrument complement. 
These articles will focus on some of the major scientific questions that the SPICA mission aims to address; more details about the mission and instruments can be found in \citet{roelfsema2017}. 

\section{Introduction}
\label{s:intro}

The cycling of matter between stars and gas is one of the drivers of galaxy evolution, along with black hole accretion and interaction with the surroundings \citep{dave2012}.
It is becoming established that the initial conditions for star formation in galaxies include large scale webs of filamentary structures that eventually form the smaller scale seeds that give birth to individual stars in molecular clouds \citep{putman2012}. 
How these processes on local galaxy scales link to a web of interstellar matter taking shape as different phases on larger scales in galaxies has yet to become clear \citep{andre2014}. 
Linking the total star formation rate (SFR) to the processes governing it in nearby galaxies is a key topic for the next decade; such detailed studies will provide a reference sample to calibrate and inform the observations of star formation across cosmic time in the coming datasets with massive samples of galaxies at all epochs.

One key finding from observations in the last $\sim$10 years is that the rate at which stars form varies widely between galaxies \citep{kennicutt2012}. 
While it is natural to expect that the SFRs of galaxies scale roughly with their stellar masses, from dwarf galaxies to normal galaxies, it is more surprising and interesting that the specific SFRs (i.e., SFRs per unit stellar mass) increase substantially from normal galaxies to starburst systems \citep{kennicutt2012,boquien2015}.
Many factors may contribute to this variation, especially morphology/galaxy type and environmental effects such as mergers and the inflow of material from the surrounding intergalactic medium.
Metallicity may also play a role, as observations show that dwarf galaxies have higher sSFRs than spirals in the local Universe \citep{brinchmann2004} and simulations show various metallicity effects on the gas/dust and H/\hh\ ratios \citep{gnedin2010,krumholz2012}. Disentangling these effects requires sensitive observations of SFRs and gas masses for a large variety of galaxies.

The interstellar media of galaxies consist of several phases of matter (ionized, neutral atomic, and molecular gas, and dust) which are actively interacting with each other and with the surrounding intergalactic medium. 
The molecular phase hosts the bulk of the star formation, but its total mass is not well constrained since the common tracer of \hh, CO, is subject to large uncertainties ({\citealt{wilson1995}; \citealt{bolatto2013}, and references therein).
In addition, observations of \fsl{C}{ii}{158}\ line emission \citep{madden1997,planck2011,pineda2013} and gamma rays \citep{grenier2005} indicate that a large fraction of \hh\ is completely invisible in CO, a fraction which increases with Galactocentric radius \citep{langer2014}.
Dust measurements also indicate substantial masses of `CO-dark gas' in low metallicity environments \citep{leroy2011}, although with caveats related to our knowledge of the dust properties {\citep{romanduval2014}. 
Unraveling the interdependence of the SFR, the properties of the multiphase gas and dust reservoirs, morphology/galaxy type and local conditions (radiation field, gas density, dynamical and chemical processes...) is crucial to our understanding of how galaxies nurture star formation and evolve over cosmic time.

A key ingredient of galaxies is their dust content, which may vary by orders of magnitude. 
The \textit{Spitzer} and \textit{Herschel} missions, together with AKARI and WISE, have provided spectral coverage of the dust emission of numerous nearby galaxies, out to the submillimeter regime \citep[e.g.,][]{smith2007,auld2012,dale2012,ciesla2014,remy-ruyer2015}. 
These data have led to major revisions of the dust models for the Large Magellanic Cloud \citep{galliano2011,gordon2014,chastenet2017} and the Galaxy \citep{ysard2015,planck2016}. 
However, our knowledge of the grain properties is still very limited and therefore the inferred dust masses for nearby and more distant galaxies are crude and potentially biased and inaccurate if Galactic dust model assumptions are made.
We still lack direct constraints on the dust composition in galaxies, because these properties have been estimated with a small number of broadband filters. 
The next level of understanding interstellar dust is to constrain the mineralogy of the bulk of the dust mass, dominating its emission spectrum.
While ISO opened the field of astromineralogy \citep[e.g.,][]{waters1996}, \textit{Herschel} 
showed how the dust content of galaxies changes with metallicity \citep{remy-ruyer2014,zhukovska2014}. 
\textit{Herschel}, however, was sensitivity-limited when it came to the low surface brightness galaxies as well as diffuse regions within galaxies. 
These include the low metallicity outer regions of (spiral) galaxies and the very low metallicity dwarf galaxies.
To date, only a handful of galaxies have been spectroscopically studied in the very low surface brightness regime where the evolution of the metals with respect to the gas and dust is highly non-linear \citep{draine2007,galliano2008b,galametz2011}. 
The very low mass, low metallicity galaxies seem to have much less dust than expected from their metallicity, which is a regime that offers us a window into the conditions in the early Universe \citep{herrera-camus2012,remy-ruyer2014}.

In summary, despite recent advances in interpreting the observations we have to date, several key questions remain open, in particular:
(i) What local and global processes determine the SFRs of galaxies?
(ii) What are the main interstellar matter reservoirs of galaxies and how do they interact?
(iii) How does the dust content of galaxies depend on the ISM phases within galaxies, the metallicity, SFR and history, and other galactic properties?

Mid- and far-infrared (IR) spectroscopy at high sensitivity is essential to take these issues a decisive step forward. 
Such spectra allow us to jointly characterize \textit{all} phases of interstellar matter (ionized, neutral and molecular gas, and dust) more robustly than can be done in any other wavelength regime, because of the many available lines and dust spectral features. 
The high linear resolution and mass sensitivity achievable in nearby galaxies (including the Milky Way) allow us to probe specific mechanisms and interpret the integrated measurements that are obtained in galaxies at higher redshift, which puts them in a more accurate physical context.

\begin{figure*}[tb]
\makebox[\textwidth][c]{\includegraphics[width=1.1\textwidth,angle=0]{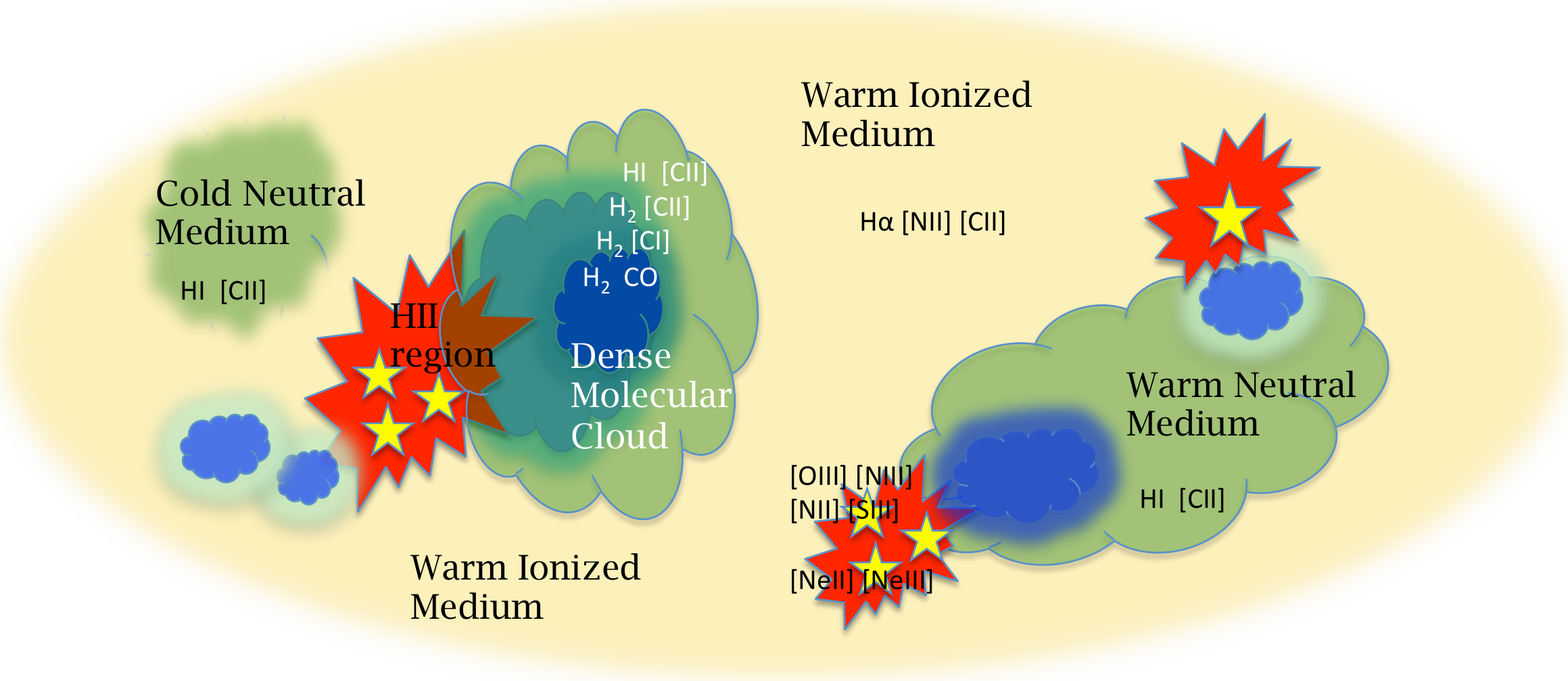}}
\caption{Schematic picture of the origin of fine-structure and molecular line emission in the mid- and far-infrared ranges from the various ISM components of galaxies.}
\label{f:lines}
\end{figure*} 

In October 2016, the \textit{Space Infrared Telescope for Cosmology and Astrophysics} (SPICA)\footnote{\tt http://www.spica-mission.org/} has been proposed to ESA by a European-Japanese-American consortium as the fifth medium-sized (M5) mission of the Cosmic Vision program. 
The basic concept of SPICA is a large space telescope with an actively cooled mirror to achieve extreme sensitivity in the mid- to far-IR wavelength ranges. 
Three instruments are foreseen: the SPICA Mid-infrared Instrument (SMI), the SpicA FAR Infrared Instrument (SAFARI), and a polarimetric imager (POL).
The design of the SPICA telescope and its instruments was initially described in \citet{swinyard2009}, and updated by \citet{nakagawa2014} and \citet{sibthorpe2016}.
The detailed design and capabilities of SPICA are described by \citet{roelfsema2017}; here we just summarize its main features.

With a 2.5\,m telescope cooled to $<$8\,K, SPICA is expected to reach 5$\sigma$ noise levels of $\sim$5$\times$10$^{-20}$\,W\,m$^{-2}$ in terms of line sensitivity in 1\,h of integration.
This sensitivity is comparable to that of the James Webb Space Telescope (JWST) and the Atacama Large Millimeter/submillimeter Array (ALMA), as shown in \citet{roelfsema2017}. 
At 34\,\mic, \spicas\ diffraction-limited angular resolution of 3.4$''$ corresponds to 170\,pc at 10\,Mpc. 
Although the critical mid- to far-IR range between JWST and ALMA was explored before by ISO, \textit{Spitzer}, and \textit{Herschel}, the 10$^2$--10$^4$ times better sensitivity of SPICA enables an entirely new approach to currently outstanding science questions.
At any given distance, SPICA will increase the accessible range in luminosities by a factor of $\sim$100. Specifically, this allows low-luminosity dwarf galaxies to be detected well beyond the Local Group, and average-luminosity spiral galaxies beyond the local universe, which has not been the case thus far.

This paper describes the progress that can be made with SPICA in our understanding of the interstellar media of nearby galaxies.
Companion papers by \citet{spinoglio2017}, \citet{gonzalez-alfonso2017}, \citet{fernandez-ontiveros2017}, \citet{gruppioni2017}, and \citet{kaneda2017}   
discuss the ability of SPICA 
to disentangle starburst nuclei from AGN, 
to study galactic winds and outflows,
to probe the chemical evolution of galaxies,
to carry out deep wide-field surveys of galaxies,
and to probe the dust evolution of galaxies.
The science case for the POL instrument is described by Andr\'e et al. (in prep), including topics in extragalactic polarimetry.
The present paper is organized as follows: 
Section~\ref{s:fs-lines} gives a brief outline of the main diagnostic spectral features in the mid- and far-IR.
Section~\ref{s:sfr} presents how SPICA will help us understand star formation processes in galaxies.
Section~\ref{s:gas} discusses the ways in which mid- and far-IR spectral lines may be used to understand the gas content of the ISM of galaxies,
while Section~\ref{s:dust} discusses the role of SPICA to understand their dust content.
Section~\ref{s:reqs} discusses the main requirements for the telescope and its instruments and outlines a possible observing program,
and Section~\ref{s:concl} concludes the paper.

\section{The power of mid- and far-infrared spectroscopy}
\label{s:fs-lines}

\begin{table*}[t]
\caption{Key diagnostic features in the mid- and far-infrared ranges.}
\label{t:far-ir}
\begin{tabular}{cccc}
\hline \hline
\noalign{\smallskip}
Species & Wavelength (\mic) & Goal \\
\noalign{\smallskip}
\hline \hline
\noalign{\smallskip}
\multicolumn{3}{c}{Ionized gas tracers} \\
\noalign{\smallskip}
\hline
\noalign{\smallskip}
O$^{++}$  & 88, 52 & shocks, ionization source \\ 
N$^+$ & 122, 205 & diffuse ionized gas, star formation rate \\ 
N$^{++}$  & 57 & hardness of radiation field \\ 
S$^{++}$, Fe$^+$  & 18, 33; 18, 26, 35 &  shocks \\ 
O$^{3+}$, Ne$^{4+}$ & 26; 14, 24 & active nucleus \\ 
Ne$^+$, Ne$^{++}$  & 12.8, 15.6, 36.0 & gas temperature \\ 
\noalign{\smallskip}
\hline
\noalign{\smallskip}
\multicolumn{3}{c}{Neutral atomic gas tracers} \\
\noalign{\smallskip}
\hline
\noalign{\smallskip}
C$^+$  & 158 & star formation rate \\ 
O & 63, 145 & UV irradiation, shocks \\ 
Si$^+$  & 35 & UV irradiation, shocks \\
\noalign{\smallskip}
\hline
\noalign{\smallskip}
\multicolumn{3}{c}{Molecular gas tracers} \\
\noalign{\smallskip}
\hline
\noalign{\smallskip}
HD  & 112, 56 & cool neutral gas \\ 
OH  & 53, 79, 84, 119, 163 & galactic winds \\ 
high-$J$ CO  & various & energetic irradiation \\
\hho\              & various & shocks \\ 
\hh\  S(0)-S(2) & 12.3, 17, 28 & warm neutral gas \\ 
HCN, HNC  & 14--15 & dense neutral gas \\ 
\coo, \cchh, \hho\ gas  & 14--15 & warm neutral gas \\ 
C$_6$H$_6$ & 14.9  & organic chemistry \\
HC$_3$N & 15.6  & organic chemistry \\
C$_6$H$_2$ & 16.1  & organic chemistry \\
C$_4$H$_2$ & 15.9  & organic chemistry \\
\noalign{\smallskip}
\hline
\noalign{\smallskip}
\multicolumn{3}{c}{Dust and ice tracers} \\
\noalign{\smallskip}
\hline
\noalign{\smallskip}
Forsterite Mg$_2$SiO$_4$ & 23, 33, 69 & dust temperature, Fe content; crystallinity \\
Enstatite MgSiO$_3$          & 28, 37, 43 & grain shape, size, composition; crystallinity \\
Calcite CaCO$_3$ & 92.6 & dust processing \\
\hho\ ice & 44, 62 & dust processing \\
MgS, graphite        & 30       & dust composition \\
Fullerenes C$_{60}$ & 17.4, 18.9  & organic chemistry \\
CO$_2$ ice & 15.2  & thermal history of ice \\
PAH (C-C-C complex) & 16--18  & PAH size \\
SiO$_2$, FeO & 20 & SN dust \\
\hline\hline
\end{tabular}
\end{table*} 

The mid- to far-IR spectral range hosts a large suite of atomic, molecular and dust features covering a wide range of excitation, density and metallicity, directly tracing the physical conditions in the nuclei, disks, and halos of galaxies (Figure~\ref{f:lines}). 
Table~\ref{t:far-ir} lists the key diagnostic features, arranged by the main ISM phase that they trace, although some features (such as [\ion{C}{ii}] 158\,\mic) trace more than one phase. 
For further information on these features, such as ionization potentials and critical densities, see the CLOUDY database\footnote{\tt http://www.pa.uky.edu/$\sim$peter/atomic/} for atomic lines, and the LAMDA database \citep{schoeier2005}\footnote{\tt http://home.strw.leidenuniv.nl/$\sim$moldata/} for molecular lines.
Fine structure lines from low-ionization species (e.g. Ne$^+$, S$^{++}$, O$^{++}$) probe \ion{H}{ii} regions around hot young stars, providing a measure of the ionizing photon flux (i.e., stellar type), and the density of the gas \citep{abel2005}. 
Lines from highly ionized species (e.g. O$^{3+}$, Ne$^{4+}$) trace the presence of energetic photons emitted from AGN, providing a measure of the accretion rate \citep{spinoglio1992}. 
Photo-dissociation regions (PDRs), the interaction zones between young stars and their parent molecular clouds, can be studied via the strong [\ion{C}{ii}], [\ion{O}{i}] and [\ion{Si}{ii}] lines and the emission from small dust grains and PAHs \citep{kaufman1999}. 
The major coolants of the diffuse warm gas (e.g. [\ion{N}{ii}] 122, 205\,\mic) in galaxies are also in the far-IR, giving us a complete picture of the warm ISM.

The rest-frame mid- and far-IR is also home to pure rotational \hh, HD and OH lines (including their ground state lines), mid- to high-$J$ CO lines, and many \hho\ lines, including the ortho-\hho\ ground state line at 179\,\mic. 
The same wavelength range hosts numerous unique dust features from minerals such as silicates and carbonates that probe evolution from pristine to processed dust (e.g. by aqueous alteration), as well as \coo\ and water ice, and molecules like acetylene (\cchh) and fullerenes, up to polycyclic aromatic hydrocarbons (PAHs).

The combination of lines in the mid- and far-IR spectra of galaxies is especially powerful as a diagnostic of physical conditions and processes in their interstellar gas. 
In particular, \fsl{C}{ii}{158}\ is often advocated as tracer of SFR and/or gas reservoir \citep{delooze2014,herrera-camus2015}, except in extreme cases such as AGN and (U)LIRGs due to the `far-IR line deficit' \citep{gracia-carpio2011,diaz-santos2013}.
In normal star-forming galaxies, both ionized and neutral gas contribute to the \fsl{C}{ii}{158}\ emission. 
To disentangle the two, the ratio of the [\ion{N}{ii}] 122 and 205 \mic\ lines may be used to estimate the density of ionized gas, so that the absolute \fsl{N}{ii}{205}\ line intensity gives an estimate of the fraction of \ion{C}{ii} emission arising from ionized gas \citep{oberst2011,hughes2014}. 
Alternatively, the ionized gas density may be estimated from the ratio of the [\ion{O}{iii}] 52 and 88 \mic, and the [\ion{S}{iii}] 18 and 33 \mic\ lines \citep{rubin1994}. 
To trace very high densities, the [\ion{Ne}{iii}] 15.6 and 36 \mic\ lines may be used.
With the gas density in hand, line ratios from adjacent ionization stages, especially [\ion{N}{ii}]/[\ion{N}{iii}], probe the hardness of the ambient UV radiation field \citep{rubin1985}. 
The ratio of the \fsl{N}{iii}{57}\ and \fsl{O}{iii}{52}\ lines is a good measure of the N/O abundance ratio, which is an indicator of the degree of stellar processing of the ISM through gas-phase metallicity \citep{rubin1988,nagao2011,pereira-santaella2017}.
The [\ion{N}{ii}] emission at 205 \mic\ has also been advocated as a good tracer of the SFR \citep{zhao2013,wu2015,hughes2016,herrera-camus2016}, as well as the combined [\ion{Ne}{ii}] + [\ion{Ne}{iii}] line emission \citep{ho2007}.

Mid- and far-IR spectra are also powerful diagnostics of neutral gas conditions, especially the [\ion{C}{ii}] and [\ion{O}{i}] lines as major cooling channels. 
The ratio of [\ion{C}{ii}] to [\ion{O}{i}] and the ratio of the two [\ion{O}{i}] lines both vary with gas density and incident UV flux, but in different ways \citep{kaufman1999}, so that SPICA can disentangle the two. 
At low metallicity, the [\ion{Si}{ii}] line is also a major coolant, potentially as bright as [\ion{C}{ii}] \citep{santoro2006}, due partly to the higher ISM temperature at low metallicity, and especially due to the low dust/gas mass ratio, with much less silicon locked into dust grains.
The HD 1--0 and 2--1 lines at 112 and 56 \mic\ are useful tracers of CO-dark gas, since their intensity ratio is a probe of gas temperature, and their absolute flux measures cool molecular gas mass. 
Given recent detections in Galactic objects, these lines are expected to be detectable as far out as $\sim$10~Mpc (see \S~\ref{ss:hd}).
In addition, the \hh\ S(0) line at 28\,\mic\ is a unique probe of warm molecular gas, which can be better detected by SPICA than by JWST, due to its cooled mirror.
The ratio of the \hh\ 17 and 28 \mic\ lines is an alternative way to measure the gas temperature \citep{stacey2010} for phases too diffuse to emit in CO. 
In nearby galaxies, the CO ladder is useful to distinguish shocks from radiative heating of the gas \citep{mashian2015}, especially the high-$J$ lines at wavelengths shorter than $\sim$200\,\mic.

In summary, high spectral sensitivity in the mid- and far-IR will enable a decisive step forward in characterizing ISM conditions in galaxies, by providing accurate estimates of the dust temperature, the gas density, and the ionizing UV radiation field (Section~\ref{s:fs-lines}; Table~\ref{t:far-ir}).
Analysis tools are available in the form of thermochemical and photoionization models such as the \citet{kaufman1999} code, the Meudon code \citep{lepetit2006}\footnote{\tt https://ism.obspm.fr/}, \textsc{Kosma}-$\tau$ \citep{roellig2007}, and \textsc{Cloudy} \citep{ferland2013}\footnote{\tt http://www.nublado.org/}, which are calibrated on Milky Way PDRs. 
Using such models, multi-line observations can be used to constrain the filling factors of the various ISM phases, as was already demonstrated for bright nearby galaxies \citep{gracia-carpio2011,cormier2012}.

\section{Star formation in galaxies: processes and rates}
\label{s:sfr}

\begin{figure}
\begin{center}
\includegraphics[angle=0,width=0.5\textwidth]{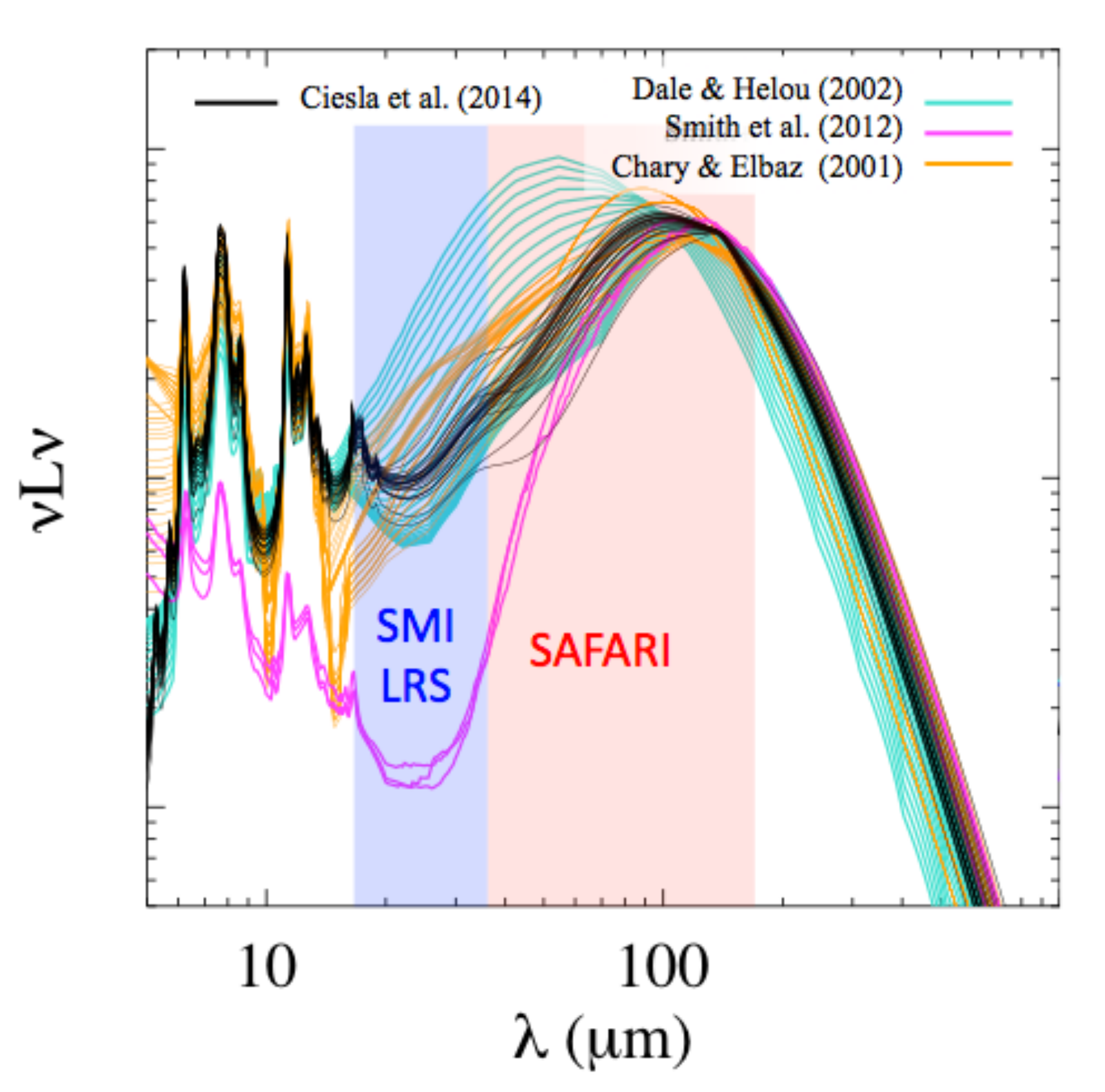}
\caption{Average observed SED from the \textit{Herschel} Reference Survey (HRS), normalized to 1\,\lsol. Superposed are model SEDs for typical dust emission templates widely used to derive the $L_{\rm IR}$ and therefore the obscured SFR, selected in the same range of $L_{\rm IR}$. Multiple lines of the same color are different models from the same library. The shape of the SED is seen to vary greatly between libraries, especially in the SPICA wavelength range.}
\label{dustmodels}
\end{center}
\end{figure}

It has recently become clear that most galaxies evolve through secular processes that involve relatively smooth accretion of metal-poor material (through minor mergers or directly from the cosmic web), star formation, and excretion of metal-enriched gas and dust into the interstellar, circumgalactic, and intergalactic medium.
The processes that regulate this cycling of matter are major drivers of galaxy evolution, and the relative importance and efficiencies of different physical processes (large-scale gas infall, outflows, AGN and stellar feedback, etc.) are not yet well characterized. 
Galaxies exhibit a wide range of star formation efficiencies, defined as the fraction of the gas mass that is going into stars. 
The availability of fresh gas and the self-regulation of star formation within galaxies determine many of their properties \citep[e.g.,][]{leroy2008,saintonge2011,alatalo2013}.
In the Milky Way, stars form inside dense cloud condensations, often filamentary, where the local SFR is directly related to the fraction of gravitationally unstable gas \citep[e.g.,][]{heiderman2010,lada2012,andre2014}. 
The idea that clouds are fully self-gravitating once they are molecular is based on the observation that the dense gas fraction (probed by the HCN/CO ratio) is remarkably constant in clouds \citep[e.g.,][]{dobbs2014}.
This picture may not hold for all galaxies: in M51 for instance, \citet{bigiel2016} find that HCN/CO is an increasing function of the local surface mass density and of the molecular gas fraction, and for Local Group galaxies, \citet{braine2017} find variations of HCN/CO with metallicity and SFR.

Compression and cooling of the ISM plays a fundamental role in setting the global and local conditions needed for star formation, while feedback provides energy that heats the gas and dust reservoirs.
However, the trigger of star formation and onset of gravitational collapse, which may determine the Initial Mass Function \citep{andre2014}, are not yet well understood.
To understand the initial stages of star formation, the process of energy dissipation of turbulent gas and the formation of filamentary structures needs to be studied in detail.

Mid- and far-IR spectroscopy affords a unique opportunity to observe major cooling and other diagnostic lines of the ISM (Figure~\ref{f:lines}); see e.g. \citet{malhotra2001}.
With its outstanding sensitivity to spectral lines, SPICA will enable detailed study of thousands of galaxies.
High spectral-line sensitivity enables measurements of the key elements that regulate star formation in galaxies:  gas temperature, density, and local radiation field conditions, both near the star formation sites and in galactic fountains and outflows.
Mid- and far-IR spectra also probe the dust content of galaxies (Table~\ref{t:far-ir}), following its evolution between the atomic and molecular gas phases and its properties in galaxy outflows that may feed the large dust reservoirs observed in the circumgalactic medium \citep{menard2010,peek2015}.

Mid- to far-IR spectroscopy over a wide wavelength range is also a great probe of Galactic ISM physics, in particular to disentangle the relative importance of radiative heating (i.e., PDRs) versus shocks. 
In the classic picture, the [\ion{O}{i}]/[\ion{C}{ii}] ratio is $>$10 if shocks dominate, and $<$10 if PDRs dominate \citep{hollenbach1989}.
However, in observing galaxies as a whole, PDRs may be dominating, while when resolving galaxies, shock-heated lines may dominate the diffuse part \citep{lee2016}.
Access to a variety of fine structure and molecular lines at a range of IR wavelengths is essential to disentangle these contributions.

\subsection{The star formation rates of galaxies}
\label{ss:sfrates}

Bright far-IR fine-structure lines can be used as star-formation rate measures at high redshift, once they are locally calibrated. 
Although local observations exist \citep[e.g.,][]{delooze2014,herrera-camus2015}, the exploration of parameter space and particularly the influence of galaxy properties (e.g., metallicity, presence of AGN, dust properties) is far from understood. 
A particularly prominent example is [\ion{C}{ii}],  which has potential as a star formation indicator and as a probe of conditions for early stages of galaxy evolution with mm-wave instruments such as ALMA. 
Herschel has clarified what drives the gas cooling, especially the [\ion{C}{ii}] emission from PDRs, in normal galaxies \citep{smith2017} and starbursts \citep{diaz-santos2013}, but these studies were restricted to bright regions in normal galaxies, or luminosity-weighted averages of $\sim$kpc scales in starbursts.
The much higher brightness sensitivity of SPICA will allow us to probe significantly fainter regions in the ISM of resolved galaxies, both normal and starbursting. 
The typical [\ion{C}{ii}] limit for galaxy surveys with Herschel was a few $10^{38}$ erg/s/kpc$^2$ ($\sim$10$^{-8}$ W/m$^2$) which is $\sim$100$\times$ higher than a short SPICA integration. 
SPICA will detect low-density gas in [\ion{C}{ii}], providing a measurement of the [\ion{C}{ii}] fraction from diffuse gas, potentially very high according to \citet{heiles1994}, which has not been done before.  
Furthermore, SPICA will detect distant objects with normal properties, not just LIRGs which often have a mixture of emission and absorption in [\ion{C}{ii}].  
SPICA will also detect outer disk material, also impossible up to now, which will help characterize the radial variation in ISM properties.

Fine-structure line tracers of SFR can also be used to investigate the correlation between SFR and metallicity.
Using the far-IR to estimate both SFR and metallicity eliminates biases due to dust extinction.
Such studies have application at high redshift, where submillimetre-wave telescopes like ALMA can access redshifted far-IR transitions \citep{swinbank2012,debreuck2014,riechers2014,aravena2016}.
Finally, accurate SPICA measurements of star formation rates will be an excellent complement to the stellar mass estimates provided for massive samples of galaxies with the coming near-IR capabilities of JWST, Euclid and WFIRST, substantially improving the measurements of the `star formation main sequence' \citep{elbaz2011}.

Besides lines from \ion{H}{ii} regions such as [\ion{Ne}{ii}] 12.8\,\mic\ and [\ion{N}{ii}] 122\,\mic, \textit{SPICA} can provide accurate measurements for the SFR in nearby galaxies based on the shape of the SED in the mid- to far-IR range. 
Such estimates are usually based either on monochromatic fluxes \citep[e.g.][]{cal00} or on total IR luminosities ($L_{\rm IR}$) derived from fitting of SED templates. 
The SFRs associated with the $L_{\rm IR}$ obtained from these libraries are provided by standard calibrations \citep[e.g.][]{kennicutt2012}, which rely either on observations of nearby galaxies \citep[e.g.][]{cha01,dal02} or on modelling of the dust emission \mbox{\citep[e.g.][]{draine2007li}}. 
Although \textit{Spitzer} and \textit{Herschel} allowed to build new templates and test them at high redshift \citep[e.g.][]{elbaz2011,magn11,magn14,mag12}, these libraries are still very generic, based on limited SED sampling, and show a large dispersion (see Fig.\,\ref{dustmodels}). 
Of particular interest is the $24$--$60$\,\mic\ range, where predictions by current templates differ by factors up to 10. 
However, it is still unclear whether the large scatter in the $24$--$60$\,\mic\ range can be ascribed only to uncertainties in the way templates are built, or if it is due to intrinsic variations in the physical properties of galaxies. 
This range is particularly sensitive to the relative contribution of photo-dissociation regions and the diffuse interstellar component to the dust heating \citep{draine2007li,draine2007,ciesla2014}. 
As shown in Figure~\ref{dustmodels}, \textit{SPICA} can drastically improve the SFR estimates by measuring the spectrum of the dust emission in the $24$--$60$\,\mic\ range.

\subsection{Structure of the ISM in nearby galaxies}
\label{ss:lg}

\begin{figure}[tb]
\centering
\includegraphics[width=0.5\textwidth,angle=0]{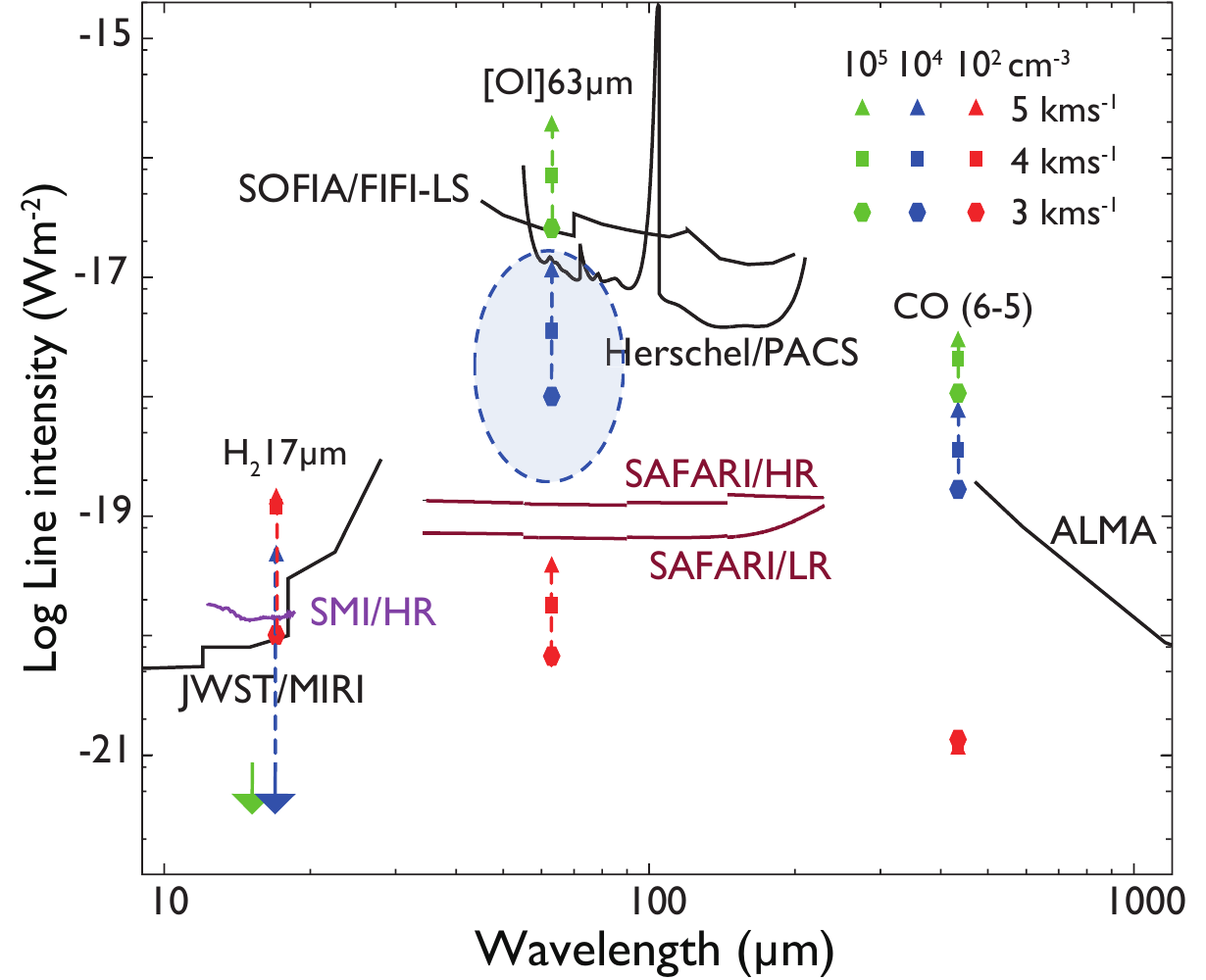}
\caption{Predicted intensities of the \hh\ S(1) 17\,\mic, [\ion{O}{i}] 63\,\mic, and CO $J$=6-5 lines for MHD shocks with different pre-shock densities and shock velocities \citep{draine1983, lesaffre2013} in comparison to SAFARI and SMI sensitivities. The blue shaded region indicates the shocks that only SPICA can trace ($10^3$--$10^4$\,cm$^{-3}$ down to 3\,km\,s$^{-1}$).}
\label{f:bontemps}
\end{figure} 

Mid- and far-IR spectroscopic mapping of Local Group galaxies holds great scientific potential because the full mix of ISM phases can be studied in detail in and around individual molecular clouds.  
At \spicas\ diffraction limit, the HD lines at 56 and 112 \mic\ can be used to map cloud complexes in the Magellanic Clouds at parsec resolution (1\,pc at 56 \mic, 2\,pc at 112 \mic), enabling a direct measure of size and molecular mass missed by CO observations (i.e. the CO-dark gas) in these environments of roughly 1/2 and 1/5 solar metallicity.  
At \spicas\ sensitivity, this can be done for clouds in many different environments in terms of radiation field and stellar surface density, in addition to the wide variety of metallicities in local galaxies.  
The HD lines are thus part of an overall study of the multiphase ISM in local galaxies which demands extreme sensitivity, which will reveal
how much gas exists in each phase, at each density, and at each temperature.
Extending such studies to large regions of local galaxies will shed light on the origin of the gas distribution over ISM phases,
in particular the roles of radiation field, local gravity, and shocks.

Although ISO and \textit{Herschel} observed the mid- and far-IR ranges, their spectroscopic sensitivity was limited so that only individual pointings or very small maps could be made of galaxies \citep[e.g.][]{mookerjea2016}.  
The $>$100$\times$ higher sensitivity of SPICA will open up new windows on the diffuse ISM and less active regions of galaxies, which is critical for assembling the complete picture of star formation in galaxies of all types -- not just the most active systems.
Furthermore, the extreme sensitivity of SPICA enables large-scale low-noise mapping to detect weak lines and measure the density, abundances, and temperatures of the \ion{H}{ii} regions, PDRs, and diffuse ionized gas throughout local galaxies, down to parsec resolution in the Magellanic Clouds.  
High spatial resolution is important because fluxes and line ratios reflect the mixture of gas found within the telescope beam.  
Throughout the Local Group, the spatial resolution of SPICA is sufficient to resolve individual molecular clouds and separate molecular clouds from PDRs and \ion{H}{ii} regions. 
It is also important to be able to isolate regions with only the warm diffuse component which sets the external pressure.  
Due to their proximity, the Magellanic Clouds in particular (49 and 69 kpc away) will be prime mapping targets.
At this distance, the diffraction limit of SPICA at 158\,\mic\ is 8--10\,pc, which is comparable to the typical sizes of molecular clouds.
These measurements were not feasible with previous IR missions due to limited sensitivity and/or wavelength coverage,  
nor are they expected with any other mission planned to date.
Such a comprehensive suite of observations will be a key step forward in our understanding of the structure and cooling of the ISM.  

In the Magellanic Clouds, one can straightforwardly map several molecular cloud complexes.  
When mapping at 3 hours per square arcminute, SPICA reaches a sensitivity such that all regions over column densities of $10^{22}$\,cm$^{-2}$ will be detected in the HD 1--0 transition for a gas temperature of 20\,K (reasonable in the LMC, temperature probably higher in SMC).
The 2--1 transition will remain undetectable, except in regions of active star formation where it will shine brightly as soon as the gas reaches 40\,K, yielding both mass and temperature independent of CO and dust emission.  
Individual single pointings can of course go much deeper in only one hour per pointing.
For the more distant Local Group dwarfs, such as WLM and NGC 6822, slightly more time per square arcminute is required but each map can be smaller, covering the region observed with ALMA by \citet{rubio2015} in WLM and the major \ion{H}{ii} regions in Hubble V and X in NGC 6822.  
The multitude of very nearby dwarf galaxies allows for a good measure of the H$_2$ column density in a variety of environments and metallicities to well characterize not only the $N$(H$_2$)/$I_{\rm CO}$ ratio but also the dust cross-section per H-atom.
The sensitivity of SPICA is such that the \ion{C}{ii}, \ion{O}{i}, \ion{O}{iii}, \ion{N}{ii}, \ion{N}{iii}, and H$_2$ lines in Table~\ref{t:far-ir}, in addition to high energy \hho\ lines, will be detected simultaneously with the HD, thus not requiring additional observing time.

Observing galaxies at a range of distances offers the possibility to connect their cloud-scale, kpc-scale and galaxy-scale properties. In particular, 
just beyond the Local Group, galaxies such as M51, M83, and M82 represent the luminosity range ($\sim L_*$) for which SPICA will provide resolved data.
They span a range of morphologies and types: M51 is a bright grand design spiral, M83 is a bright strongly barred spiral, and M82 is a small but starbursting object.  
At intermediate redshifts (i.e., $z\sim0.5$), SPICA would detect such galaxies in both spectroscopy and continuum, but not resolve them.
In such cases, having local benchmarks is key to measure whether, how, and why intermediate-redshift galaxies differ from local objects. 
Several square arcminutes of M51, M83, and M82 can be mapped with SPICA at a sensitivity similar to that of the local group dwarfs (discussed above) in order to detect HD in similar situations (to identify the differences) as well as all the other lines.  
In the local objects, SPICA can clearly separate the nuclear regions from the disk but also identify emission characteristic of a bar and separate the arm and interarm regions at a variety of radial distances.  
Studies with {\it Herschel} have shown significant differences between the nuclei of galaxies and their disks, both in the molecular abundances \citep{gonzalez-alfonso2012} and in the cosmic-ray ionization rate \citep{vdtak2016}.
Comparing the HD, \ion{C}{ii}, \ion{O}{i}, \ion{O}{iii}, \ion{N}{ii}, \ion{N}{iii}, and H$_2$ lines to local galaxies, SPICA will enable us to estimate the fraction of emission coming from the different regions.  
Where previous IR space telescopes were only able to observe the extreme starbursts ($>$10$^{12}$\,\lsol), the sensitivity of SPICA enables rather normal star-forming galaxies to be detected at intermediate redshifts, and thus to study how the conditions in these non-exceptional objects change with cosmic time.
 
\subsection{Energy dissipation from clouds to star-forming regions}
\label{ss:shocks}

Large ($\sim$kpc) scale motions in galaxies generate interstellar turbulence, and the corresponding turbulent kinetic energy is either dissipated by velocity shears (non-compressive mode; \citealt{falgarone2009}) or leads to the formation of dense structures at the origin of star formation through shocks (compressive mode; \citealt{godard2009}). 
Molecular clouds are partly supported by these turbulent, supersonic motions, and dissipation of this kinetic energy (turbulent dissipation) is therefore central to the regulation of star formation. 
The current picture is that turbulent dissipation occurs in low-velocity shocks which are closely related to the formation of filamentary structures (\citealt{padoan2001,pudritz2013}). 
The resulting warm gas has been proposed to correspond to the observed excess of mid-$J$ CO emission in nearby clouds \citep{pon2012,larson2015}. 
Subsequent formation of star-forming dense filaments may be associated with shocks in intermediate-density regions ($10^3$--$10^4$\,cm$^{-3}$), where a large fraction of kinetic energy is released through [\ion{O}{i}] 63, 145\,\mic\ and CO line emission \citep{draine1983,lesaffre2013}. 

While low-density ($\sim$10$^2$\,cm$^{-3}$) turbulent dissipation will be studied by JWST \hh\ observations, high-sensitivity spectroscopy of a wide range of far-IR lines is crucial to correctly estimate the contribution from PDRs and study shocks down to low velocities. 
The regime below 3-4 km\,s$^{-1}$ at densities of $10^3$--$10^4$ cm$^{-3}$ is well probed by the major [\ion{O}{i}] 63\,\mic\ cooling line, which traces the bulk of the energy dissipation processes to form star-forming filaments (Figure~\ref{f:bontemps}). 
While oxygen is likely partially locked up in ice at $A_V$ $>$2--3, depletion is unlikely for the shocked material considered here.
With SPICA we will be able to obtain the first global picture of the energy dissipation processes from low-density shocks to filaments and star formation in the Galactic ISM. 
This will complement ALMA low-$J$ CO observations, clearly identifying how shocks and turbulent dissipation occur, and clarifying the connection with the formation of dense filamentary structures and massive ridges of cluster-forming clumps.

\section{Interstellar matter reservoirs of galaxies}
\label{s:gas} 

Most star formation activity in galaxies occurs in and is fueled by the molecular gas reservoir, which cannot be probed directly due to its low temperature. 
Cold \hh\ is usually probed by CO observations \citep[e.g.][]{bolatto2013}, but it is becoming increasingly clear that using only CO may miss a large fraction of the total molecular gas residing in galaxies \citep{madden1997,grenier2005,planck2011}. 
In such ÒCO-dark gasÓ, the dust extinction is low and CO is photodissociated but \hh\ can survive due to its self-shielding effects. 
In our Galaxy, the mass of CO-dark gas ranges from at least 30\% to $\sim$100\% of that traced by CO, with variations depending on the metallicity gradient from the inner to the outer Galaxy \citep{langer2014}.  
For the half-solar metallicity spiral M33, \citet{gratier2017} found that CO traces only $\sim$50\% of the molecular gas, without any radial dependence.
The `missing molecular mass' problem is acute for the low-extinction ISM of low metallicity galaxies \citep{wolfire2010,nordon2016},
where vigorous star formation is often observed, but little or no cold gas reservoir is detected in the form of CO \citep[e.g.][]{schruba2012,cormier2014,cormier2017}. 
Using tracers such as HD and [\ion{C}{ii}], SPICA can uncover the total \hh\ reservoir in such galaxies where CO-based estimates may be off by orders of magnitude \citep{hunt2015,israel2015}.

\subsection{Probing cool gas with HD}
\label{ss:hd}

\begin{figure}[tb]
\centering
\includegraphics[width=7cm,angle=0]{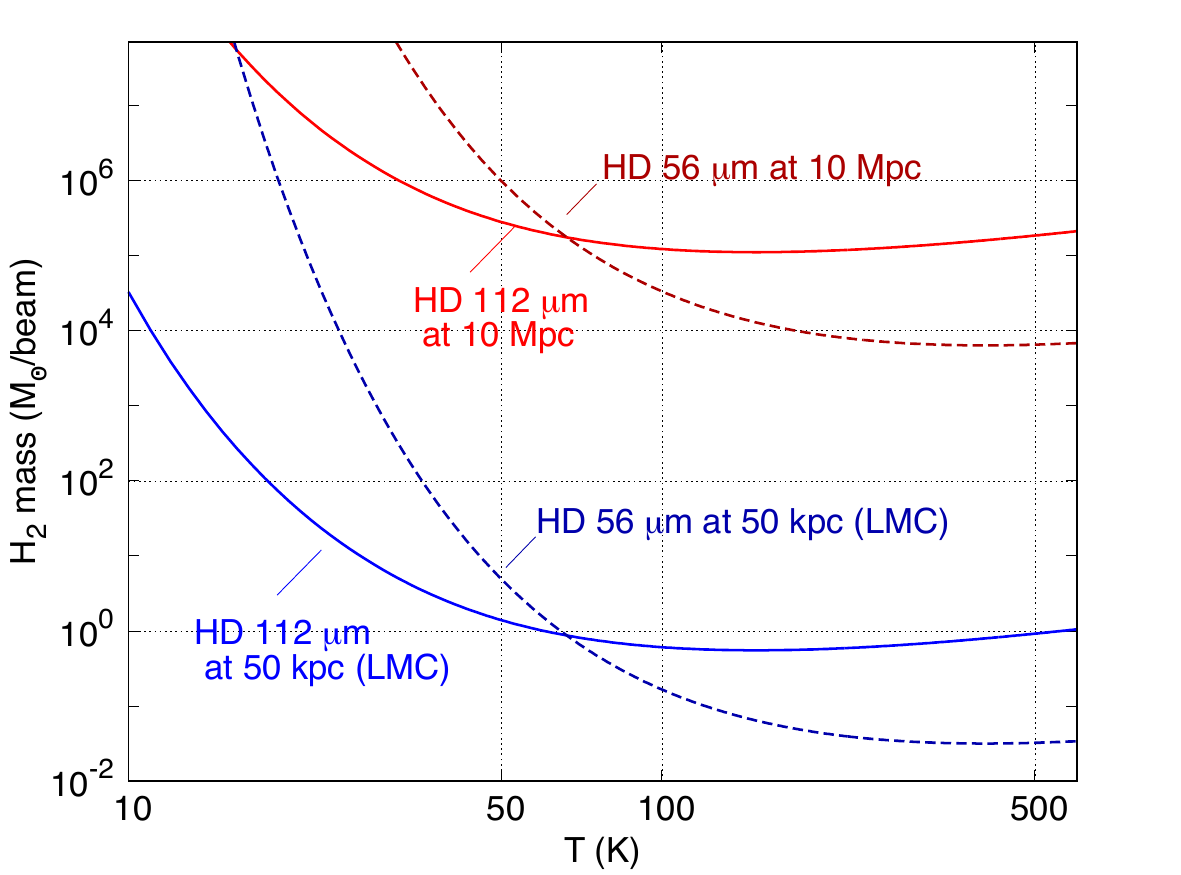}
\caption{Detectable molecular mass in HD as a function of distance and gas temperature, for a flux limit of \pow{5}{-20} W\,m$^{-2}$ (SAFARI R=300).  }
\label{f:kaneda}
\end{figure} 

The full impact of metallicity on the ISM and star formation and their interplay can only be understood via detailed studies of the different galactic ISM phases. 
One of the outstanding issues today is how vigorous star formation can exist in low metallicity environments where CO, the common tracer of the \hh\ reservoir, is difficult to detect.
The size and the nature of the gas reservoir in these systems is unknown, and hence the efficiency of the gas to cool and collapse into stars.
To detect and characterize the gas reservoirs of galaxies, the HD lines at 112 and 56\,\mic\ play a key role, as they are the most direct tracers of cool molecular gas (\hh, $T \gtsim$30--50\,K). 
Using both lines gives a direct handle on the temperature dependence of the emission \citep[e.g.][]{trapman2017}. 
The bulk of the CO-dark gas is at low extinctions ($A_V$=0.5--1.0) where gas temperatures exceed $\sim$30~K \citep{wolfire2010}, which makes HD a good probe of this gas.
The D/H ratio is well determined ($\approx$25 ppm: \citealt{linsky2006}) and the effect of astration is small for the Solar neighbourhood ($\approx$5 ppm), including the effect of infalling pristine gas from the Galactic halo. 
For dwarf galaxies with low SFRs, the effect of astration should be even smaller, but enhanced HD photodissociation may be more important.
Observations of molecular D/H ratios with ALMA may help to limit uncertainties in the HD abundance.
With ISO and \textit{Herschel}, the HD lines have been observed toward a few Galactic objects \citep{wright1999,bergin2013,mcclure2016}, but the $\sim$100 times higher sensitivity of SPICA/SAFARI will allow their detection in nearby galaxies as well (Figure~\ref{f:kaneda}), which, when combined with gas temperature estimates from HD or other molecules, will give accurate \hh\ masses. 
For example, at temperatures of 20-30 K, giant molecular clouds (GMCs) of $10^6$--$10^7$ \msol\ are detectable out to $\sim$10\,Mpc distance, which covers $\sim$100 spiral galaxies. 
In dwarf galaxies, gas temperatures may be higher, so that smaller clouds and larger distances become detectable.
High-sensitivity HD observations will thus shed light on `CO-dark' gas, because they give a tracer in which this gas is not dark anymore.

\subsection{Probing CO-dark gas with fine-structure lines}
\label{ss:c+}

\begin{figure}[tb]
\centering
\includegraphics[width=7cm,angle=0]{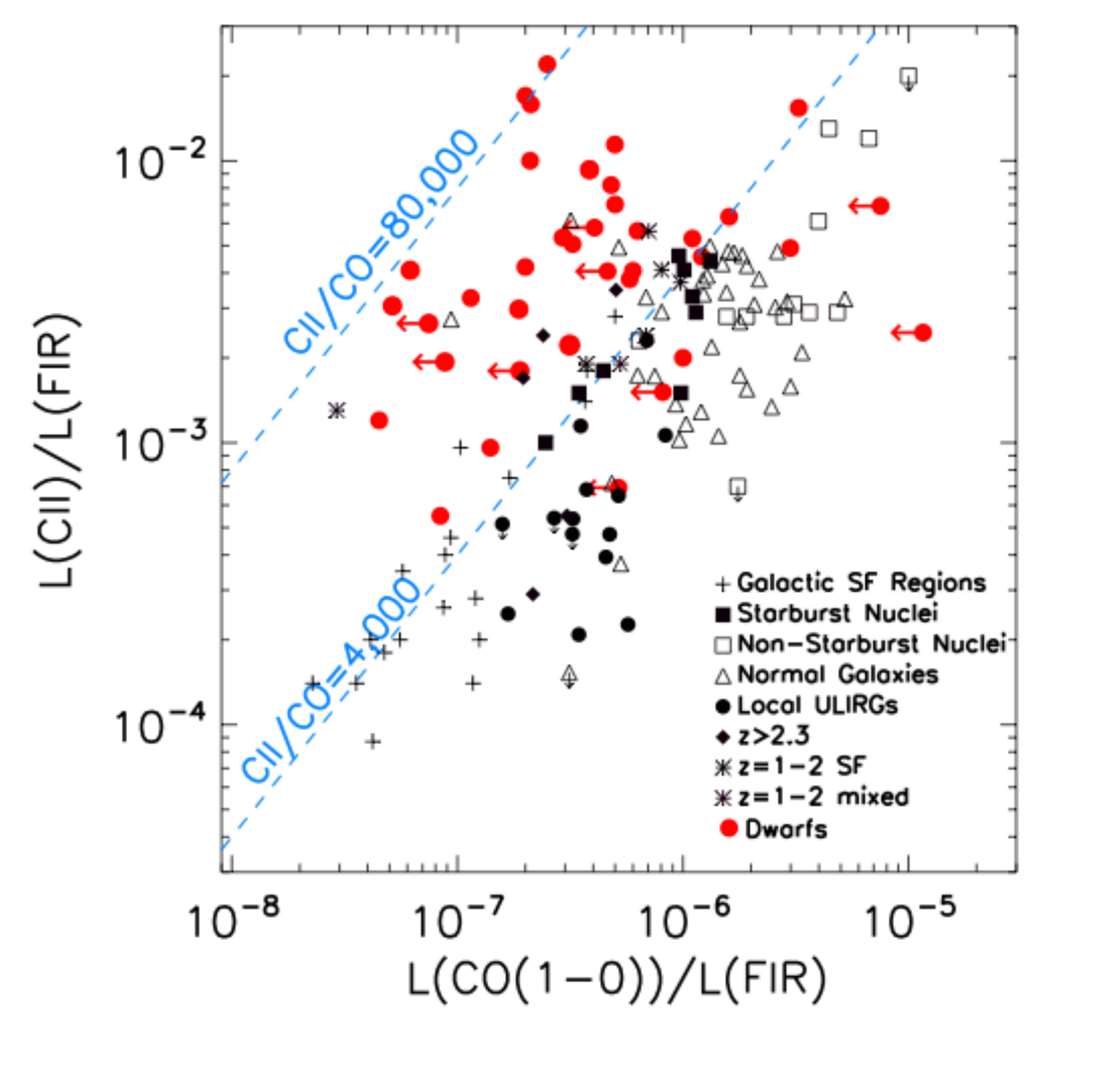}
\caption{Luminosity in [\ion{C}{ii}] versus that in CO, both normalized to the far-infrared luminosity, for galaxies with widely ranging star forming activity: quiescent galaxies, starburst galaxies, ULIRGS, high-z galaxies, and dwarf galaxies. Dashed lines are examples of constant [\ion{C}{ii}]/CO ratios. While most normal and SB galaxies show L[\ion{C}{ii}]/CO between 1000 and 4000, low-metallicity dwarf galaxies show much higher ratios, up to an order of magnitude.  From \citet{madden2016}.}
\label{f:madden}
\end{figure} 

For distant galaxies, the HD lines are too weak to detect, and fine structure lines will be key to measuring the total molecular gas reservoir. 
For the expected sensitivity of SPICA, this limit lies at $\sim$10 Mpc (Figure~\ref{f:kaneda}).
Studies with \textit{Herschel} show that the [\ion{C}{ii}] line is relatively bright in low-metallicity galaxies even though CO is faint (Figure~\ref{f:madden}), as expected for stronger radiation fields \citep{nordon2016}.
Recent ALMA observations indicate that low-metallicity galaxies contain large \cp-\hh\ zones where CO is confined to small dense clumps \citep{indebetouw2013,rubio2015,chevance2016}. 
Contributions to the \cp\ emission by atomic gas \citep{lebouteiller2017} can be calibrated with the [\ion{N}{ii}] 122, 205\,\mic\ and [\ion{O}{i}] 63, 145\,\mic\ lines.
Mid- and far-IR lines (Table~\ref{t:far-ir}), including the [\ion{C}{ii}] 158\,\mic\ line, can be used to quantify the total molecular gas reservoir in galaxies, including the CO-dark \hh\ that resides outside of the CO-emitting region. 
Comparison of HD and [\ion{C}{ii}] observations of nearby ($<$10~Mpc) galaxies will be key to calibrate the two tracers against each other.

With \spicas\  sensitivity, we will obtain a complete picture of the ISM in the lowest metallicity galaxies, including their currently elusive molecular phase. 
A comprehensive suite of the mid- and far-IR fine structure lines provided by SPICA, such as \fsl{C}{ii}{158}, \fsl{O}{i}{63}, \fsl{O}{i}{145}, \fsl{N}{ii}{122}, \fsl{O}{iii}{88}, \fsl{O}{iii}{52}, \fsl{N}{iii}{57}, \fsl{Si}{ii}{35}, and the [\ion{S}{iii}] and 33\,\mic\ lines, covers 8 to 40 eV in ionization energies and from 100 to 10$^5$ cm$^{-3}$ in gas densities. Such a wide range of properties provides a full picture of the various phases of galaxies, such as the filling factors of the dense and diffuse ionized and neutral gas, and trace their temperatures, densities, abundances, and radiation fields. A glimpse of what will be possible for many galaxies is demonstrated in \citet{cormier2012}. 

\subsection{Diffuse gas in galaxies}
\label{ss:diff}

\begin{figure}[tb]
\centering
\includegraphics[width=7cm,angle=0]{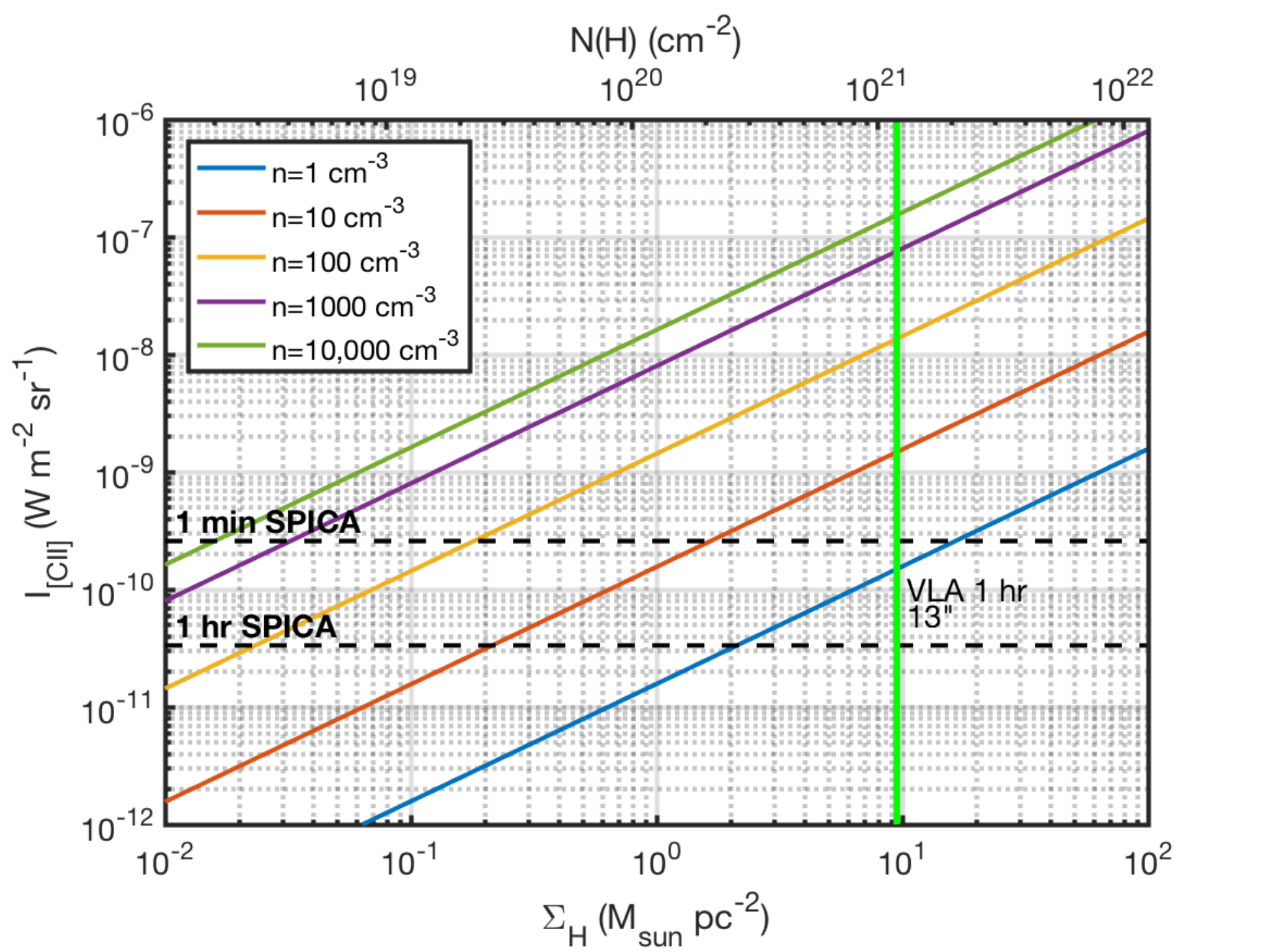}
\includegraphics[width=7cm,angle=0]{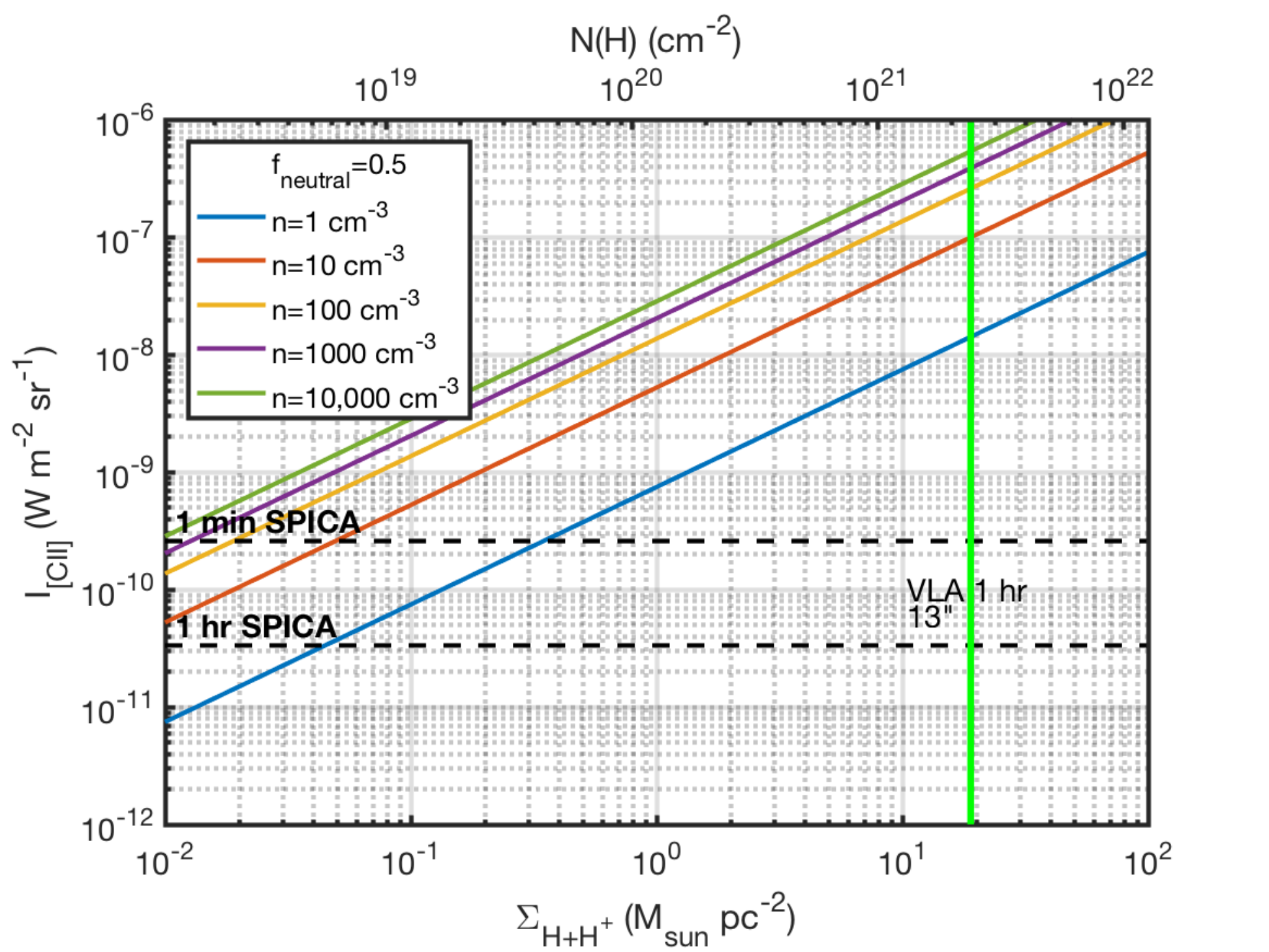}
\caption{Predicted [\ion{C}{ii}] 158\,\mic\ emission versus column and surface density (top and bottom axes) of warm diffuse atomic gas, with the sensitivities of SPICA and the VLA indicated. The top panel assumes purely neutral gas; the bottom panel assumes 50\% ionized gas. The gas temperature is assumed to be $>>$92~K, and the effect of gas density is indicated. All carbon is assumed to be in C$^+$, with a Solar carbon abundance. The width of the lines is taken to be 125\,km\,s$^{-1}$.}
\label{f:alberto}
\end{figure} 

A major challenge in understanding the ISM of nearby galaxies is imaging material in diffuse regions with low gas column densities, which requires high surface brightness sensitivity. 
Mapping [\ion{C}{ii}] emission with SPICA is ideal for this purpose, especially on scales $\gtsim$10$''$ where radio interferometers lack sensitivity.
In high-resolution mode, the longest-wavelength channel of SAFARI offers a 5$\sigma$ flux limit of \pow{5.1}{-19} W\,m$^{-2}$ for an 1$'$x1$'$ field (Roelfsema et al. 2017), which corresponds to a 1$\sigma$ surface brightness sensitivity of \pow{1.9}{-11} W\,m$^{-2}$\,sr$^{-1}$ in 15 seconds.
This sensitivity corresponds to a C$^+$ column density $N_{C^+}$ of $$ \frac{h\nu}{4\pi} A_{ij} N_{C^+} f_u$$
where $h\nu$, $f_u$ and $A_{ij}$ are the photon energy, the upper level population, and the spontaneous decay rate of the [\ion{C}{ii}] 158\,\mic\ line \citep{crawford1985}.
For a gas density well above the critical value of $\approx$4000\,cm$^{-3}$ and a kinetic temperature well above $E_{\rm up} / k_B \approx 92$\,K, this yields  $N_{C^+}$ $\sim$\pow{1.3}{13}\,\scm; 
for the cases of lower gas temperatures and/or densities, see Figure~6 of \citet{madden1993}.
Assuming a solar carbon abundance with all carbon in C$^+$, and neglecting contributions from ionized gas to the [\ion{C}{ii}] emission, the corresponding hydrogen column density is $\sim$\pow{3.8}{16}\,\scm.
Accounting for the contribution of He to the mass, this column corresponds to a surface density of $\sim$\pow{4}{-4}\,\msol\,pc$^{-2}$, which is orders of magnitude below the threshold of 1--2 \msol\,pc$^{-2}$ \citep{martin2001,schruba2011}, below which star formation is not observed to occur.
For comparison, Jansky VLA observations of the \ion{H}{i} 21\,cm line, matched to the same 13$''$ resolution, reach a column depth of \pow{1.4}{21}\,\scm, which is $\sim$10$^4$ times less deep.
The SKA is currently planned to have 3$\times$ the collecting area of the VLA, mostly on short baselines, and thus $\sim$3$\times$ the VLA sensitivity, which is still well below \spicas\ capabilities for warm gas (Figure~\ref{f:alberto}).
With this high brightness sensitivity, SAFARI will be uniquely suited to probe extraplanar material and extended UV disks \citep[e.g.][]{thilker2007}, measuring the in-situ conditions for star formation in a regime that has been completely inaccessible in the past.
Similarly, SPICA will have extraordinary potential for tracing gas ejected in galactic fountains or winds, which in nearby galaxies can be imaged using far-IR transitions, particularly the [\ion{C}{ii}] line \citep[e.g.,][]{contursi2013}.

\section{The origin and evolution of dust in galaxies}
\label{s:dust}

\begin{figure*}[tb]
\centering
\includegraphics[angle=0,width=0.8\textwidth]{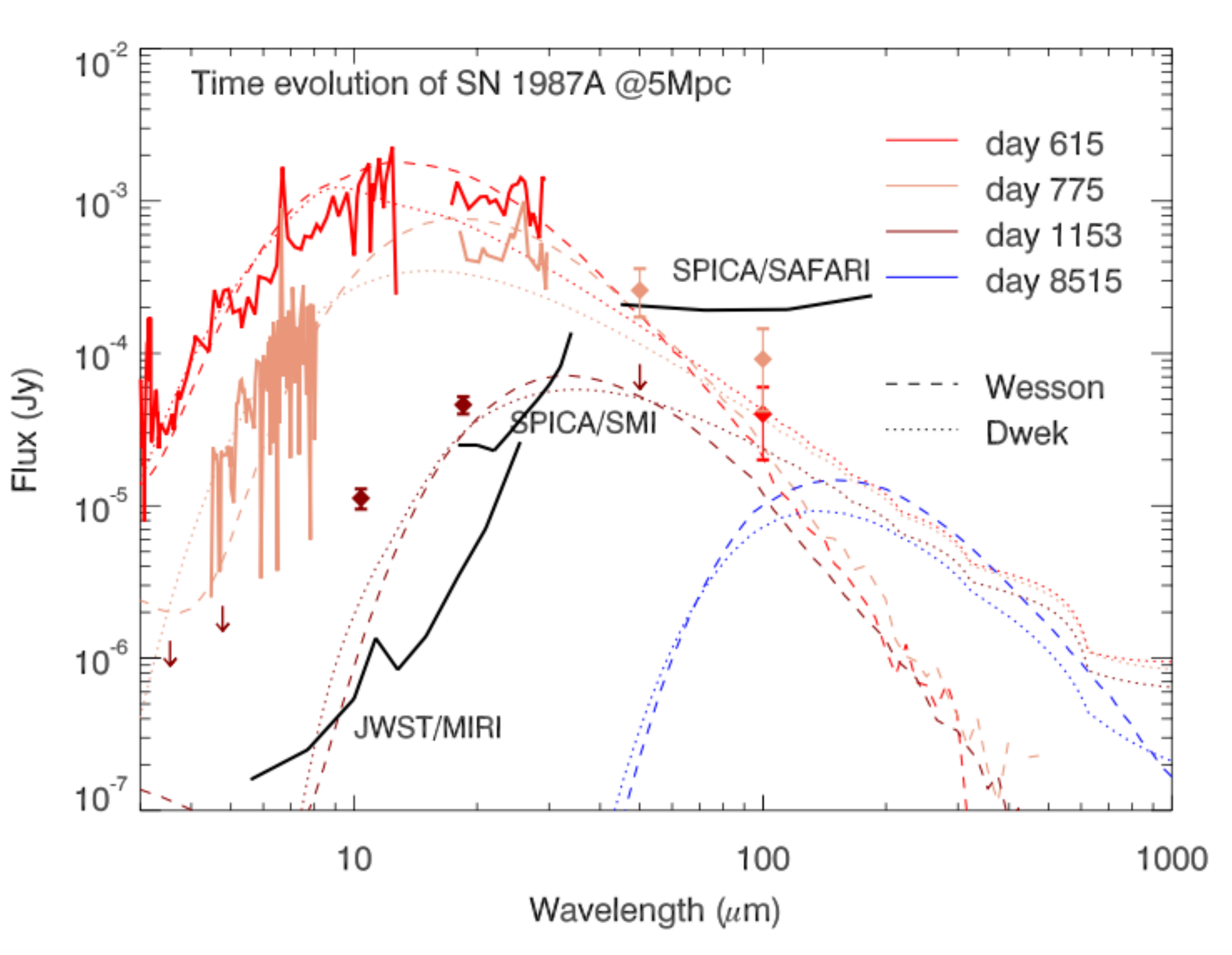}
\caption{Time evolution of the SN 1987A ejecta, scaled to 5~Mpc. 
The SED of the dust emission has peaks shifting from the mid- to the far-IR in time. 
The observed data (histograms and points with error bars) are fitted with two different types of models: (i) an initially small dust mass (10$^{-3}$\,\msol), increasing to 0.8\,\msol\ over 20 years (Wesson et al. 2014), and (ii) a large dust mass from the beginning (0.4\,\msol), with the optical depth changing in time, as the ejecta expand (Dwek \& Arendt 2015). 
Accurate photometry with SPICA/SMI has the potential to disentangle the two different models.}
\label{f:mikako}
\end{figure*} 

In normal star-forming galaxies (SFR $\sim$1\,\msol\,yr$^{-1}$), the fraction of starlight absorbed by dust and re-emitted in the far-IR is determined to be about one third on average \citep{popescu2002,skibba2011,viaene2016}. 
In actively star-forming galaxies (SFR $\gtsim$10\,\msol\,yr$^{-1}$), this fraction goes up to 90\%, and even higher in ULIRGs like Arp 220, which is evidence that IR emission from dust is the major cooling process during star formation in galaxies. 
Although dust is key to ISM energetics and regulating SFRs \citep[e.g.,][]{forbes2016}, its formation and the path to its ultimate composition remain poorly understood.
Simultaneous spectroscopy of dust and gas over a wide mid- and far-IR wavelength range will unveil and characterise the chemical and physical conditions where dust grains are formed, processed, and destroyed in unprecedented detail. 
The dust emission from galaxies peaks in the far-IR wavelength range, and complements observations of mid-IR dust and PAH features with JWST, in particular toward AGB stars.

\subsection{Dust content of nearby galaxies} 
\label{ss:mass}

In the traditional view of galaxy enrichment, star formation results in metal production through nucleosynthesis in massive stars on short ($\sim$Myr) timescales. 
The presence of a large amount of refractory elements enables condensation of a large dust mass in galaxies with strong star formation. 
Indeed, the total dust masses of galaxies generally appear to correlate with their two major properties: star-formation rate and metallicity \citep{dacunha2010}, and some high-redshift galaxies clearly were already dust-rich \citep[e.g.][]{laporte2017}.
However, recent discoveries of starbursting dwarf galaxies with little dust call into question the standard picture \citep{remy-ruyer2014,fisher2014}.
Equally interesting and somewhat hard to interpret is the recently discovered class of low-mass galaxies with a large dust mass and a high SFR, but very little UV attenuation \citep{clark2015,devis2017}.
Also, in damped Lyman $\alpha$ systems, the dust to gas mass ratio measured in absorption seems to follow a linear trend with metallicity, down to $Z$=10$^{-2}$\,\zsol, suggesting that the star formation history plays a role. 
A large systematic survey measuring the dust masses of galaxies in the local universe is needed to understand the origin of the observed relations between dust mass and star formation rate, metallicity, and galaxy type.
More modeling is needed to quantify how far SPICA can push the current limits in terms of specific parameters, in particular dust mass, SFR, and metallicity.

The primary means to measure dust masses is the submillimeter range, for which good quality data exist from the \textit{Herschel} and \textit{Planck} missions, but the main uncertainty in their interpretation is the dust emissivity. 
In the LMC \citep{galliano2011}, the far-IR grain emissivity is found to be 2-3 times higher than in the widely used prescription of \citet{draine2007li} which fits data for the Milky Way \citep{planck2014}.
Detailed panchromatic radiative transfer studies of edge-on spiral galaxies point in a similar direction \citep{baes2011,degeyter2015,mosenkov2016}. 
Getting the dust emissivity correct throughout the mid- to far-IR is essential to estimate extinctions at optical/UV wavelengths, which are key to derive stellar masses and SFRs. 
The uncertain emissivity calls into question the dust chemical composition in our Galaxy and its variation within galaxies \citep{fanciullo2015}. 
Local conditions such as gas density or dust temperature may affect the dust emissivity \citep{mennella1998,ysard2015}, as suggested by apparent dust emissivity changes across galaxies \citep{smith2012,tabatabaei2014}.  
These results demonstrate the need for more reliable measurements to constrain dust models, also for the interpretation of high-redshift observations. 
Far-IR observations are especially crucial for low-metallicity galaxies, for which the dust emission peaks shortward of 100\,\mic\ and where free-free emission outshines dust at longer (ALMA) wavelengths.

Sensitive far-IR observations with SPICA would allow us to study dust in the diffuse ISM of nearby galaxies, which is useful because it is the phase that is the closest to uniform illumination, so that radiative transfer effects (temperature mixing, etc.) can be neglected. 
Since CO-dark gas and optically thick HI are negligible for this phase, the gas mass from HI 21 cm data can be used to calibrate the dust properties. 

Understanding dust properties as a function of metallicity is an essential step toward accurately probing the dust masses of high-redshift galaxies, which tend to have metallicities well below solar. 
Dust at high redshift is a topic of great current interest \citep{watson2015,laporte2017}, but results up to now have been very uncertain. 
Attempting to go into the lowest metallicity regime, \textit{Herschel} surveys of the observed gas-to-dust (G/D) ratio in galaxies over 2 dex in metallicity range show a striking non-linear evolution, with little dust present in lower metallicity galaxies \citep{remy-ruyer2014}. 
While supernovae are clearly a significant source of heavy elements, their dust destruction rates may exceed their dust formation rates, so that supernovae may not be sufficient to account for observed dust masses at low metallicity.
Chemical evolution models require grain growth by mantle accretion which is efficiently producing dust at $Z / Z_{\odot} \gtsim 0.1$ \citep[e.g.,][]{asano2013,zhukovska2014,feldmann2015} to explain the observations. 
Lower metallicity environments require more time to accumulate metals for grain growth \citep{remy-ruyer2014,zhukovska2014}. 
To develop these models into diagnostic prescriptions, the low metallicity part of parameter space must be well sampled observationally, for which \textit{Herschel} lacked sensitivity.
Detailed models are required to establish to which depth in metallicity the extreme sensitivity of SPICA will allow us to probe.
By detecting and characterizing the dust content of extremely low metallicity galaxies for the first time, SPICA would make a link with galaxy formation in the early Universe.

\subsection{Dust formation in supernovae and AGB stars}
\label{ss:sne}

\begin{figure}[tb]
\centering
\includegraphics[width=0.5\textwidth,angle=0]{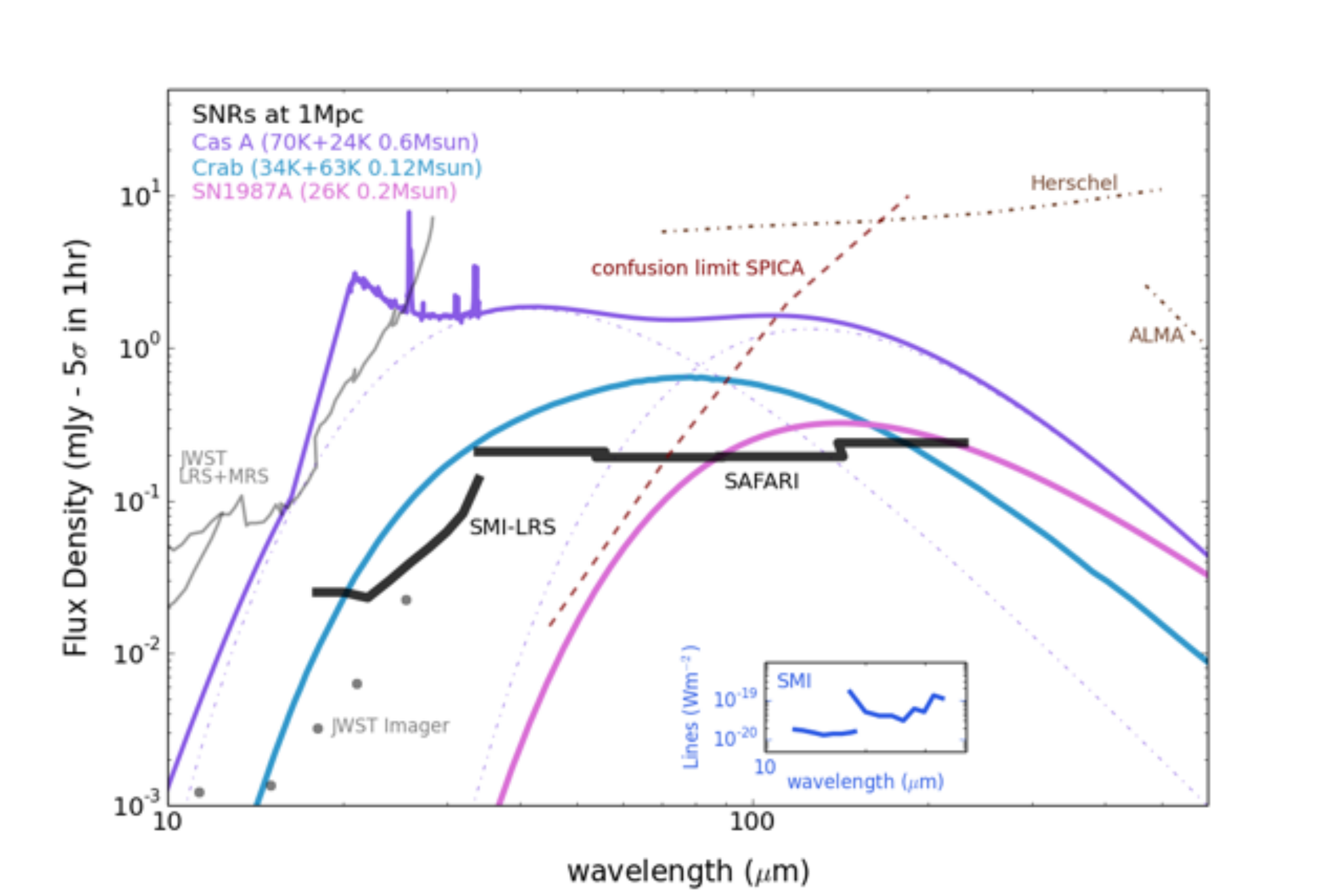}
\caption{Sensitivity comparison for SPICA SAFARI and SMI (solid black lines) compared to {\it Herschel}, ALMA and {\it JWST} in dot-dashed grey.  
The inset shows the 5-$\sigma$, 1 hour levels, line sensitivities for SPICA SMI MRS and HRS spectrometers.
The confusion limit is shown as a dashed line.}
\label{f:gomez}
\end{figure} 

Major outstanding questions concerning dust in galaxies are the locations where dust grains are formed, and how much mass is formed by each dust source. 
Two major sources of the elements that make up dust exist: AGB stars and supernovae. 
In the Milky Way, AGB stars have been identified as an important dust source, but this result does not seem to hold across other types of galaxies. 
Recent estimates of the dust mass and composition in the Magellanic Clouds suggest that AGB stars are not dominant contributors to the present day dust reservoir \citep{matsuura2009,boyer2012,kemper2013}.
Other evidence suggests that a substantial part of the dust mass must have been formed in the ISM \citep{mattsson2014,remy-ruyer2015}. 
Deep mid- and far-IR spectra of gas and dust for a wide range of local conditions in the ISM across many galaxies are required to constrain the local physical conditions (temperature and density) needed for particular dust processes to occur. 
For instance, \citet{devries2012} have shown how the central position and the width of the 69\,\mic\ forsterite feature can be used to determine the iron content and the temperature of the dust grains, even at R=300.
Also, \citet{planck2016} have found variations of dust properties with radiation field intensity, which has led to a new dust model by \citet{jones2016}.
 
As an alternative to AGB stars, there is growing evidence that the aftermaths of supernova explosions are major sites of dust formation 
\citep{rho2009,barlow2010,matsuura2011,gomez2012,delooze2017}, 
and supernovae can be a major source of dust in galaxies \citep[e.g.][]{dwek2011,rowlands2014,michalowski2015}. 
Our current understanding of dust formation in SNe is presently hampered by the small sample size, limited wavelength coverage, and limited time frame. 
While changes from circumstellar dust to ejecta have been reported for various recent supernovae \citep[e.g.,][]{gall2014}, our only supernova example of long-term dust evolution is provided by SN 1987A \citep{wesson2015,bevan2016}. 
Its dust thermal emission was initially detected at near- and mid-IR wavelengths within one year after the explosion, with an estimated dust mass of $\sim$10$^{-4}$\,\msol\ \citep{wooden1993}. 
The mid-IR emission disappeared in 2 years, while suddenly after 25 years thermal dust emission appeared in the far-IR, with an estimated mass of $\sim$0.5\,\msol\ \citep{matsuura2015}. 
Currently, we have no idea how the dust mass increased between 3 years and 25 years, leaving a big gap in our understanding of dust formation in SNe. 
\citet{wesson2015} suggested that the peak of the SED gradually shifted from short to long wavelengths over 25 years, with increasing dust mass. 
An alternative scenario is that a large mass was present from early days but emission was optically thick, and previous authors underestimated dust masses \citep{dwek2015}.
Monitoring supernovae in Local Group galaxies at far-IR wavelengths is needed to settle this issue, with a time base of at least a few years.
Besides sufficient sensitivity to detect supernova dust in external galaxies, a key requirement is the ability to resolve supernova dust from circumstellar and interstellar dust, which SPICA can only do for very local galaxies. 
Covering the full mid- to far-IR wavelength range is essential to measure the mass of warm dust produced in supernovae.
Ultimately, we hope to understand if supernovae are a net source or sink of dust.

Based on current models (Figure~\ref{f:mikako}), we predict SPICA/SMI to be able to take spectra of SNe exploding out to $\sim$40\,Mpc.
The expected number of core collapse SNe within this distance is estimated to be $\sim$40 per year, based on the observed rate in nearby galaxies \citep{smartt2009}.
This amounts to $\sim$200 SNe over the 5-year mission lifetime, of which about 5 exploding within 5~Mpc can be followed up with SAFARI. 

While detecting extragalactic supernovae in the mid- and far-IR remains a challenge even for SPICA, prospects are better for supernova remnants, due to their larger sizes.
Figure~\ref{f:gomez} shows the predicted spectral energy distributions of local supernova remnants (SNRs), Cas A ($\sim$300 yrs old), Crab Nebula ($\sim$1000 yrs old) and SN1987A at day $\sim$10,000 scaled to a distance of 1~Mpc.  
The predicted SEDs for SNRs like the Crab, Cas A, and SN\,1987A SNRs at 1\,Mpc are shown in order of top to bottom: 
the Crab Nebula ($\sim 0.116\,\rm M_{\odot}$ including the cool $T=34$ and warm 64\,K components from \citealt{gomez2012}); 
Cas A (Spitzer spectra from \citealt{rho2008}, warm and cold SN dust peaks from \citealt{delooze2017} at $\sim$23K and 70K); and 
SN1987A ($\sim$0.2\,\msol, $T=26$\,K \citealt{matsuura2011,matsuura2015}). 
The figure shows that it is possible to detect the Crab and Cas A SNRs to $\sim$1\,Mpc distance with SAFARI (at least the peak of the emission) while SN1987A at 10,000 days is more difficult.
However, SPICA will still provide crucial mid-IR detections of these objects with SMI. 
The sensitivity of this instrument (an order of magnitude better than JWST) has the potential to provide key mid-IR measurements of SNR dust.  
For example, \citet{rho2008} used \textit{Spitzer}  spectra to reveal a new 21~\mic\ dust feature as well as spectral lines from atomic species to 
reveal never-seen-before supernova dust freshly forming.
Figure~\ref{f:gomez} demonstrates that SPICA also has the potential to reveal the composition of the mid-IR-emitting dust and gas within Cas A or Crab-like SNRs in external galaxies and provide a measurement of the mass of hot, freshly formed dust.  
The mid-IR holds the key dust features unique to particular compositions, such as SiO$_2$ and FeO at 20\,\mic, and FeS at 30\,\mic.

\subsection{Grain processing and destruction}
\label{ss:cryst}

Besides dust formation, SPICA offers an opportunity to investigate the composition and evolution of interstellar dust in galaxies. 
Crystalline silicates in the warm circumstellar envelopes of evolved stars are readily detected in the 10-45 $\mu$m range, while crystalline silicates in colder ISM dust are better probed with the 69\,$\mu$m feature (Table~\ref{t:far-ir}). 
In principle, the crystalline fraction of silicates in AGB stars can be measured using diagnostic tools, such as those proposed by \citet{kemper2001} and \citet{devries2010}. 
However, a systematic determination of the crystalline fraction of AGB stars observed with spectrographs on board of ISO and \textit{Spitzer}  is still missing, and determinations of crystalline fractions remain anecdotal. 
Estimates in common use place the crystallinity of the AGB silicates around 10\%, however some extremely crystalline sources have been reported \citep{jiang2013}. 
\citet{jones2012} has performed the most complete inventory into the presence of crystalline silicate features in AGB stars and Red Supergiants, in the Milky Way and both Magellanic Clouds, demonstrating a correlation with mass-loss rate; however, they did not measure the crystalline fractions. 

In the ISM, cosmic-ray processing amorphitizes these crystalline silicates, and the ratio between the crystalline fractions of circumstellar and interstellar silicates places constraints on the fluence of low-energy cosmic rays in galaxies \citep{kemper2011}.  
Cosmic ray amorphization is most efficient for heavy ions with energies around 30-60 keV. 
The cosmic ray flux at these relatively low energies cannot be measured directly, as the solar wind blocks such particles from entering our Solar System, which makes the measurement of silicate amorphization an interesting alternative to constrain the low-energy cosmic ray flux. 
Furthermore, the shape of the 69\,$\mu$m crystalline silicate feature changes with the composition of silicates. 
Magnesium-rich silicates have a peak at short wavelength, and shifts towards longer wavelengths are indicative of a high iron content \citep{blommaert2014}. 
Far-IR spectra can thus potentially determine the composition of silicate dust, which is a long standing issue.

Interstellar shocks, caused by supernova blast winds and active galactic nuclei, may destroy existing ISM dust \citep[e.g.][]{jones1994}, 
but the efficiency of dust destruction is poorly measured by observations, especially for the carbonaceous dust component \citep{jones2011}. 
To constrain these theories, determination of elemental abundances across the ISM of galaxies on $\sim$kpc scales are required, including super bubbles and galactic outflows, which are triggered by SNe. 
Dust destruction would result in releasing refractory elements into their gas phase, thus elemental abundances will vary across galaxies \citep{lebouteiller2008}. 

Observations with SPICA may contribute to our understanding of interstellar dust destruction in two ways.
The dust continuum SED is a good probe of the dust size distribution in galaxies, which is indicative of dust destruction processes.
In addition, mid- to far-IR spectroscopy provides us with the gas-phase metal abundance, which gives the amount of heavy elements returning to the gas phase by shock processing of dust.
An example of the first approach exists for small areas on the sky \citep{arendt2010}, and \spicas\ sensitivity would enable such studies across entire galaxies, as done in the SMC by \citet{sandstrom2010}, although further modeling efforts are needed.
The second approach is interesting as a direct probe of the dust composition, as outlined in \citet{fernandez-ontiveros2017}.
At present, we have Fe and Si abundances from \textit{Spitzer}  observations \citep{okada2008}, but a good reference is lacking.
With SPICA, we can obtain the N abundance from [\ion{N}{iii}] and [\ion{N}{ii}] lines as a proxy for H, and also the O and C abundances, which are the major dust components.
These results would give a complete picture of the cycling of metals from gas to dust and back.

\section{Mission requirements and observing strategy}
\label{s:reqs}

By taking full advantage of the diagnostic power of gas-phase lines and solid-state features in the mid- to far-IR wavelength range (Table~\ref{t:far-ir}), a spectroscopic survey of nearby galaxies with SPICA would initiate a revolution in galaxy ISM science. 
A key feature of SPICA is that its spectrometers (SMI and SAFARI) provide full-band coverage with a single measurement.
The SPICA mission is able to cover the entire known ranges in luminosity, metallicity, SFR, and morphological type, in a volume- or magnitude-limited sample of compact ($\sim$10$''$) galaxies at distances of 50--100 Mpc, so that the objects fit into a single pointing. 
Covering each of these parameters in $\sim$5 bins would imply a sample size of about $\sim$3,000. 
In addition, to resolve the inner workings of galaxies spatially, SPICA should observe 10--100 pointings in hundreds of nearby ($\sim$10 Mpc) galaxies to unravel the contributions of various components (nuclei, disks, spiral arms, etc) to the mid- and far-IR spectrum. 
Progress will be especially outstanding for dwarf galaxies, both star-forming and irregular, where the EUCLID mission will increase the number of known objects by orders of magnitude.
Equally exciting would be to probe the interstellar media of the recently discovered 'ultra-diffuse' galaxies \citep{vandokkum2015}.

Whereas the current SPICA concept fulfills many of the above science requirements, several further improvements are desirable.
Obviously, using the largest possible telescope mirror is vital to maximize both the sensitivity and the angular resolution.
In the context of nearby galaxies, the higher sensitivity would especially help to probe further out into the halos of galaxies and trace the interaction with their surroundings.
The smaller beam size would especially be helpful for the polarimetric imaging of star-forming regions with the POL instrument (Andr\'e et al., in prep.), but also to resolve supernova dust from the galaxy background.
To fit within the spacecraft, elliptical mirror shapes may have to be considered.
The alternative of cooling the mirror below 8~K is lower priority, as this increases the sensitivity mainly at the longest wavelengths ($>$200\,\mic).
To probe ISM physics, more useful would be to increase the number of pixels on the sky, so that the distribution of interstellar gas and dust could be mapped faster, especially for very nearby systems (\S\ref{ss:lg}).
Extending the spectral coverage beyond 210\,\mic\ would allow observation of the C$^+$ line beyond $\sim$1000\,Mpc.
Finally, extending the mission lifetime beyond 5~years would improve the monitoring timebase of supernovae and supernova remnants.

\section{Conclusions}
\label{s:concl}

The SPICA mission will be a leap forward in our understanding of the baryon cycle of dust and gas in galaxies and the formation of stars. 
Herschel was only able to measure the 2-3 brightest far-IR lines in typical galaxies, leaving large uncertainties in global conditions resulting from model degeneracies.   
With SPICA we will measure dozens of mid- and far-IR spectral features (gas-phase and dust), providing a full characterization of the interstellar reservoirs and star-forming processes within galaxies.  
In addition, SPICA will be sensitive to all physical components of galaxies (disks, halos, spiral arms, etc.) and show us how they interact. 
But most importantly, SPICA will observe statistically significant samples of thousands of galaxies, which will give us an unbiased view of these building blocks of our Universe.

While these results are valuable in themselves, they also impact other studies. 
In particular, SPICA observations of nearby galaxies will be essential to understand the mid- and far-IR spectra of high-redshift galaxies \citep{spinoglio2017}. 
Vice versa, SPICA observations of the ISM in local galaxies will inform studies of how such processes depend on the properties of the galaxies,
and spatially resolved observations will characterize how these processes depend on the local environment. 

\begin{acknowledgements}
This paper is dedicated to the memory of Bruce Swinyard, a major force behind the initial SPICA project, who died on 22 May 2015 at the age of 52. 
We remember him as ISO-LWS calibration scientist, \textit{Herschel}-SPIRE instrument scientist, initial European PI of SPICA, and initial design lead of the SAFARI instrument. 

\bigskip

We appreciate the helpful coments by an anonymous referee which have improved the paper.
One of us (HG) acknowledges support from ERC Consolidator Grant CosmicDust. 

\end{acknowledgements}

\begin{appendix}
\section*{Affiliations}
\affil{$^1$ SRON Netherlands Institute for Space Research, Landleven 12, 9747 AD Groningen, The Netherlands}
\affil{$^2$ Kapteyn Astronomical Institute, University of Groningen, The Netherlands}
\affil{$^3$ CEA Saclay, France}
\affil{$^4$ California Institute of Technology, USA}
\affil{$^5$ University of Gent, Belgium}
\affil{$^6$ Department of Physical Sciences, The Open University, Milton Keynes, UK}
\affil{$^7$ University of Maryland, USA}
\affil{$^8$ Universit\'e de Bordeaux, France}
\affil{$^9$ Universit\'e de Strasbourg, CNRS, Observatoire Astronomique de Strasbourg, UMR 7550, F-67000 Strasbourg, France }
\affil{$^{10}$ Imperial College London, UK}
\affil{$^{11}$ ZAH/ITA Heidelberg, Germany}
\affil{$^{12}$ Instituto de Astrof\'isica de Canarias (IAC), Tenerife, Spain}
\affil{$^{13}$ Universidad de La Laguna, Dpto. Astrof\'isica, Tenerife, Spain}
\affil{$^{14}$ IAPS/INAF, Rome, Italy}
\affil{$^{15}$ IRAP Toulouse, France}
\affil{$^{16}$ Cardiff University, UK}
\affil{$^{17}$ Universidad de Alcal\'a, Departamento de F\'isica y Matem\'aticas, Alcal\'a de Henares, Madrid, Spain}
\affil{$^{18}$ NRC Herzberg Astronomy and Astrophysics,Victoria, Canada}
\affil{$^{19}$ Department of Physics and Astronomy, University of Victoria, Canada}
\affil{$^{20}$ Institut d'Astrophysique Spatiale, Paris, France}
\affil{$^{21}$ Nagoya University, Japan}
\affil{$^{22}$ Academica Sinica, Institute of Astronomy \& Astrophysics, Taipei 10617, Taiwan}
\affil{$^{23}$ University College London, UK}
\affil{$^{24}$ Department of Space Astronomy \& Astrophysics, ISAS/JAXA, Japan}
\affil{$^{25}$ University of Tokyo, Japan}
\affil{$^{26}$ Universidad Complutense, Madrid, Spain}
\end{appendix}

\bibliographystyle{aa}
\bibliography{ism,metaldust}

\end{document}